\documentclass[fleqn,10pt]{wlscirep}
\usepackage[utf8]{inputenc}
\usepackage[T1]{fontenc}
\usepackage{braket}
\usepackage{dsfont}
\usepackage{graphicx,epsfig}
\usepackage{amsmath,amssymb,bm}
\usepackage[normalem]{ulem}

\title{Experimental demonstration of scalable quantum key distribution over a thousand kilometers}

\usepackage{tabularx}
\usepackage{makecell}

\author[1]{A.\,Aliev}
\author[1]{V.\,Statiev}
\author[1]{I.\,Zarubin}
\author[1]{N.\,Kirsanov}
\author[1]{D.\,Strizhak}
\author[1]{A.\,Bezruchenko}
\author[1]{A.\,Osicheva}
\author[1]{A.\,Smirnov}
\author[1]{M.\,Yarovikov}
\author[1]{A.\,Kodukhov}
\author[1]{V.\,Pastushenko}
\author[1]{M.\,Pflitsch}
\author[1]{V.\,Vinokur}

\affil[1]{Terra Quantum AG, St. Gallen, 9000, Switzerland}
\affil[*]{vv@terraquantum.swiss}

\begin{abstract}
Secure communication over long distances is one of the major problems of modern informatics. 
Classical transmissions are recognized to be vulnerable to quantum computer attacks.
Remarkably, the same quantum mechanics that engenders quantum computers offers guaranteed protection against such attacks via quantum key distribution (QKD). 
Yet, long-distance transmission is problematic since the essential signal decay in optical channels occurs at a distance of about a hundred kilometers.
We propose to resolve this problem by a QKD protocol, further referred to as the Terra Quantum QKD protocol (TQ-QKD protocol).
In our protocol, we use semiclassical pulses containing enough photons for random bit encoding and exploiting erbium amplifiers to retranslate photon pulses and, at the same time, ensuring that at the chosen pulse intensity only a few photons could go outside the channel even at distances of about a hundred meters.
As a result, an eavesdropper will not be able to efficiently utilize the lost part of the signal.
The central component of the TQ-QKD protocol is the end-to-end loss control of the fiber-optic communication line since optical losses can in principle be used by the eavesdropper to obtain the transmitted information. 
However, our control precision is such that if the degree of the leak is below the detectable level, then the leaking states are quantum since they contain only a few photons. 
Therefore, available to the eavesdropper parts of the bit encoding states representing `0' and `1' are nearly indistinguishable. Our work presents the experimental demonstration of the TQ-QKD protocol allowing quantum key distribution over 1079 kilometers. 
Further refining the quality of the scheme's components will expand the attainable transmission distances.
This paves the way for creating a secure global QKD network in the upcoming years.
\end{abstract}
\begin{document}

\flushbottom
\maketitle
\thispagestyle{empty}

\section{Introduction}
The TQ-QKD protocol\,\cite{new_theory} offers the resolution of the problem of secure long-distance communications by providing two remote legitimate users with quantum protected keys that can be utilized for further secure information transmission.
The emergence of Shor's algorithm created a threat of breaking currently standardized public key algorithms (e.g. the RSA\,\cite{rsa}, ECC\,\cite{ecc}, and DSA\,\cite{dsa}).
Although Shor's algorithm can only be executed on massive quantum computers that as yet do not exist, this threat must not be ignored.
Fortunately, the same quantum physics brings in a possibility for fully secure communication methods such as quantum secure direct communication\,\cite{qsdc:1, qsdc:2, qsdc:3, qsdc:4, qsdc:5, qsdc:6, secure classical repeaters} and quantum key distribution\,\cite{qkd:1, qkd:2, qkd:3, qkd:4, qkd:5, qkd:6, qkd:7}. 
The latter, which is the focus of the current study, enables communicating parties to generate a shared random secret key known only to them which then can be used to encrypt and decrypt messages.
It enables communicating parties to generate a shared random secret key known only to them, which then can be used to encrypt and decrypt messages.
A unique property of quantum key distribution providing its security relies on the foundations of quantum mechanics: the ability of the communicating parties to detect the presence of any third party trying to gain knowledge of the key.
This results from the fundamental quantum mechanics principle, the fact that measuring disturbs the measured quantum states.
A third party trying to eavesdrop on the key inevitably creates detectable anomalies.
By careful analysis of transmitted quantum states, a communication system detects eavesdropping and, if necessary, takes measures to fully secure the distribution.

Most of the existing QKD applications are curtailed by channel losses that result in the exponential decay of the signal with the distance. 
For repeaterless point-to-point QKD, the secret key generation rates cannot surpass fundamental Pirandola-Laurenza-Ottaviani-Banchi (PLOB) bound\,\cite{plob}.

In the framework of repeaterless point-to-point QKD, up-to-date record secret key generation rates\,\cite{compare_qkd_mdi, compare_qkd_bb84} do not exceed several bits per second at distances of 400\,km.
To overcome the PLOB bound, one can introduce trusted nodes\,\cite{trusted_nodes:1, trusted_nodes:2, trusted_nodes:3} where local secret keys are produced for each QKD link between nodes and stored in the nodes.
This implies re-coding of the quantum information into a fully classical one at these trusted nodes and completely eliminates quantum protection of an overall protocol.
The way to preserve quantumness is to use quantum repeaters.
Ideally, quantum repeaters\,\cite{quantum_repeaters:1, quantum_repeaters:2, quantum_repeaters:3, quantum_repeaters:4, quantum_repeaters:5, quantum_repeaters:6, quantum_repeaters:7, quantum_repeaters:8, quantum_repeaters:9, quantum_repeaters:10, quantum_repeaters:11, quantum_repeaters:12, quantum_repeaters:13, quantum_repeaters:14, quantum_repeaters:15, quantum_repeaters:16} would have been expected to decontaminate and forward quantum signals without directly measuring or cloning them.
However, such idealized quantum repeaters remain unavailable for existing technologies. 
Twin-Field QKD protocol\,\cite{lucamarini, compare_qkd_tf, compare_qkd_tf_1002} which is based on sending quantum states from Alice and Bob to an intermediate point represents another way to overcome PLOB bound and effectively double the maximum achievable distance.

\begin{figure}[h!]
    \centering
    {\includegraphics[width = 0.6\linewidth]{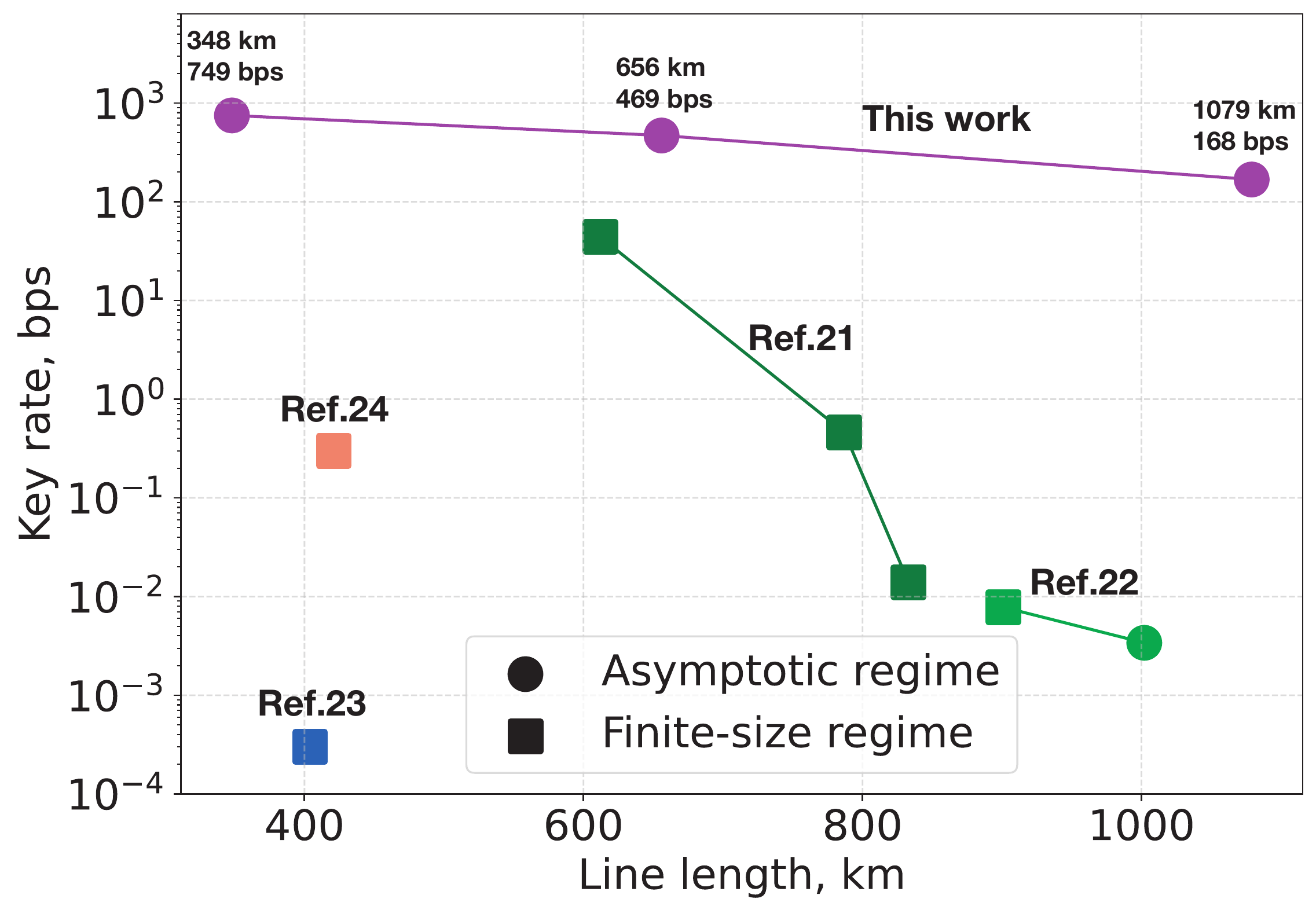}}
    \caption{
    \textbf{
    Landscape of long-distance ($\gtrsim400$\,km) QKD realizations\,\cite{compare_qkd_tf, compare_qkd_tf_1002,compare_qkd_mdi, compare_qkd_bb84}.}  
    The plot shows dependence of the secret key distribution rate on the length of the optical fiber over which the key is distributed. The secret key that accounts for the finite-size statistics is marked by squares, and the secret key in the asymptotic regime is marked by circles.}
    \label{comparing}
\end{figure}

To show the current situation in this research area, Fig\,\ref{comparing} summarizes some of the previous realizations of long-distance QKD including the current record distance\,\cite{compare_qkd_tf_1002}.
Our work presents the realization of the TQ-QKD protocol eliminating the PLOB bound by using quantum thermodynamic restrictions and quantum mechanics-based loss control in the optical channel.
The implementation of the secure long-distance transmission line is based on using the Erbium-Doped Fiber Amplifiers (EDFA) of our own Terra Quantum construction arranged every 50\,km which have enabled the practical realization of the 
fiber-optic communication system over 1079\,km.
In Figure\,\ref{comparing}, we provide a set of experimental data points with different optical line lengths to demonstrate the robustness and scalability of our system.
Note that here different classes of device dependency are shown, as well as different conditions concerning finite-size or asymptotic scenario.
Below, we present the results of measurements for our best result with the optical line length of 1079 km.

\bigskip\section{TQ-QKD experimental setup}\label{QKD_section}

\begin{figure}[h!]
    \centering
    {\includegraphics[width = 1.0\linewidth]{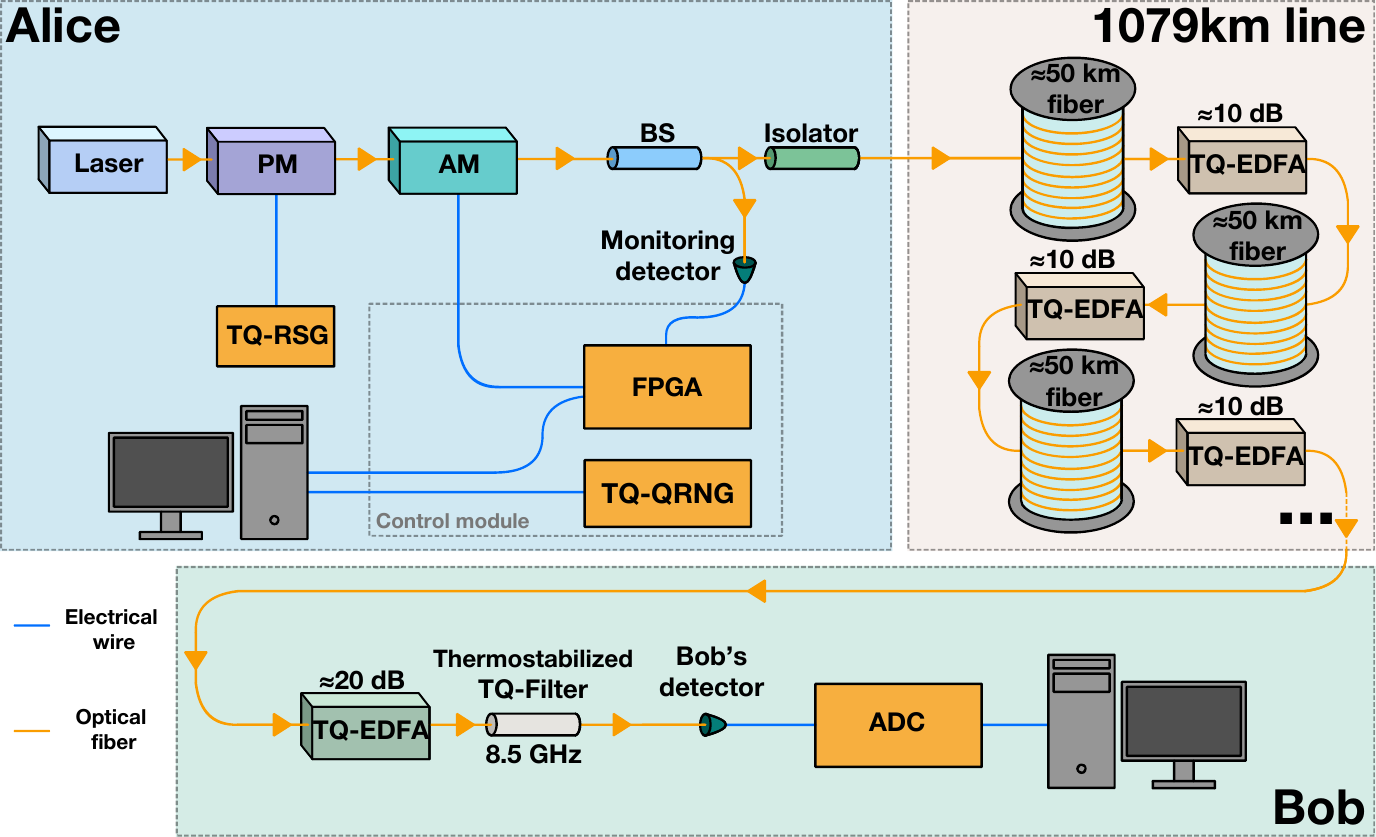}}
    \caption{
    \textbf{Experimental setup of the QKD system}. 
    Using the original Terra Quantum-made quantum random number generator (TQ-QRNG), Alice creates the bit sequence. 
    Field programmable gate array (FPGA) transforms bit sequences into voltage pulses of 2.5\,nano\,sec each and sends them to Alice's amplitude modulator (AM). 
    The phase modulation is implemented by a phase modulator (PM) which receives the random signal from the Terra Quantum-made random signal generator (TQ-RSG). 
    Alice's source spot comprises the laser, PM, and AM connected via the polarization-maintaining optical fiber. 
    Alice's AM comprises the Mach–Zehnder interferometer, a device determining the relative phase shift variations between two collimated beams derived by splitting the light from a single source and then modulating the intensity of the light beam from the laser source. 
    The beam splitter (BS) sends the part of the signal coming from the AM back to Alice for controlling and proper tuning of the modulation process. 
    The main part of the fiber optic communication line comprises a sequence of approximately 50\,km-long fiber channels, each ending with the TQ-EDFA having an amplification factor of about 10\,dB. 
    The receiving user, Bob, uses the TQ-EDFA with an amplification factor of 20\,dB to enhance the distinguishability of the incoming signal and the optical thermal-stable Terra Quantum-made filter (TQ-Filter) to extract the information wavelength from the noise-distorted signal. 
    The filtered signal is measured by the photodetector and is finally treated by Bob's analog-to-digital converter.}
    \label{QKDscheme}
\end{figure}

The Terra Quantum QKD system realizing the secure 
key distribution between the legitimate users, Alice and Bob, comprises the single mode fiber segments connected via Terra Quantum-made EDFAs (TQ-EDFA). 
Here we report on the setup with the amplifiers of approximately 10\,dB amplification coefficient, and with G.652.D single mode fiber segments of approximately 50 km length. 
The experimental setup of the implemented TQ-QKD is shown in Fig.\ref{QKDscheme}. 
At Alice's laboratory, the signals that are to be sent via the fiber optical channel are formed from the laser pulses by the amplitude modulator (AM). 
Before arriving at AM, generated pulses pass through the phase modulator (PM), which shifts the phase of each pulse randomly. Phase randomization\,\cite{phase_rand, phase_rand2, phase_rand3, phase_rand4} decreases the effectiveness of the eavesdropper's attacks without affecting the probability distributions of Bob's measurement results. 
The phase shift of a particular pulse is controlled by the Terra Quantum-made random signal generator (RSG) using the avalanche breakdown described in Refs.\,\cite{qrng:1,qrng:2}.
Exploiting the quantum interference effect, the AM modulates the intensity of the optical signal. 
The phase difference between the interfering components is set by the voltage coming to the radio frequency (RF) port. 
Further, an additional direct current (DC) port is used for the bias point shifting. 
The high-frequency electrical signal is created using a field programmable gate array (FPGA) and an electrical amplifier.
The random bits are generated using the TQ-QRNG that has been certified by the Federal Institute of Metrology METAS (Test Report No 116-05151).
Our TQ-QRNG uses the random time interval between the registrations of the photons generated by a weak LED light source via a single-photon detector as an entropy source. The hashed time intervals constitute the final random number sequences.
The FPGA board sets in the constant offset voltage for the bias control getting the information about the present power level coming from the AM via the monitoring detector. 
The light radiation generated by the laser has a 1530.33\,nm wavelength. 
This wavelength is chosen since it corresponds to the peak of the EDFA amplification spectrum, see Fig.\,\ref{amplif_spectrum}.

At Bob's spot, Alice's optical signals are first amplified by the 20\,dB TQ-EDFA, then go through the narrow-band TQ-made optical filter with the 8.5\,GHz bandwidth, see Supplementary Information (SI), Note\,5. Since this filter is the fiber Bragg grating (FBG), its bandwidth is temperature sensitive. To stabilize the temperature of the narrow bandwidth filter, the Terra Quantum team developed a specific thermostat. Being filtered from the natural amplifier's noise referred to as an amplified spontaneous emission (ASE), Alice's signal arrives at Bob's detector. Then the voltage coming from the detector is digitized by the analog-to-digital converter, and processed automatically by Bob's computer.
The models of equipment used can be found in SI, Note\,3.

\bigskip\section{TQ-QKD protocol}\label{protocol_section}

\begin{figure}[h!]
    \centering
    {\includegraphics[width = 1.0\linewidth]{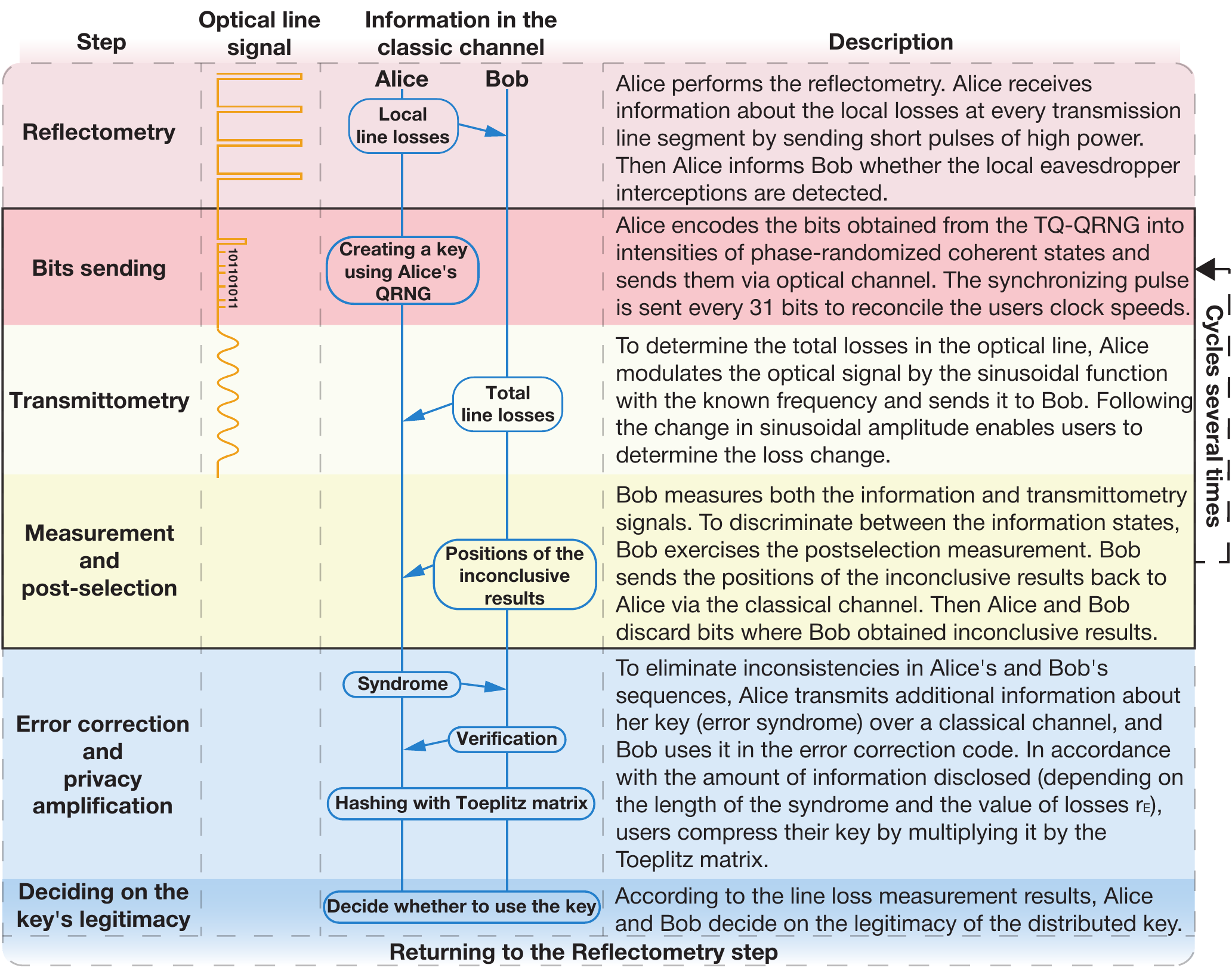}}
    \caption{\textbf{Basic structure of the TQ-QKD protocol.}
    The figure shows the main stages of the protocol and their description together with an illustration of the information transmitted through the classical and optical channels. The short strong probing pulses for performing reflectometry, the sync pulses together with the `0' and `1' intensity coding states, and the sinusoidal signal used for transmittometry sent over the fiber-optic line are shown in orange.  
     After measuring the optic signal, only the authorized
classical channel, shown in blue, remains. 
    The information transmission operation between the legitimate users is depicted by the arrows. 
    }
\label{protocol_block_scheme}
\end{figure}

The structure of the TQ-QKD protocol is shown in Fig.\,\ref{protocol_block_scheme}. Before starting the secret communication, legitimate users must make sure that there is no eavesdropper intercepting  the transmission line.
To that end, ideally, they have to execute the loss control procedure using the optical time domain reflectometry (OTDR) technique. 
Corresponding devices commonly used for testing the integrity of fiber lines not containing amplifiers are described below in Section\,\ref{REFL_section}.
We develop a special OTDR technique suitable for our TQ-QKD protocol which we describe in detail in a forthcoming publication. 
The reflectometry procedure is repeated with a certain frequency.
It allows us to determine the reference value of losses in the optical line, which slowly drift with time. 
This drift is one of the main problems that creates the need for periodic repetition of the reflectometry procedure. 
As an additional degree of control, the OTDR can also be used before the start of the protocol to check the unique backscattered signatures of all fiber spans that are planned to be used in the line for QKD\,\cite{reflectokey}. As each signature demonstrates physical unclonable function behavior, one can pre-measure the backscattered radiation variations at the time when the optical fiber is stored, e.g., in the fiber spools (and cannot be possessed by Eve), and then check the line in the process and after installation.
The actions of `bits sending' and `transmittometry' occur between each pair of consecutive acts of reflectometry.
The reflectometry and transmittometry facilitate continuous loss control in parallel with the bit distribution. 

At the beginning of the key generation process, Alice encodes the logical bits into phase-randomized coherent states. The pulses corresponding to the different bits have different average photon numbers and random phases. At his side, Bob carries out the measurements of the energies of the received states and exercises the subsequent classical post-selection of the results. The measurements can be formalized using the projective operators
\begin{equation}
    \hat{E}_0=\sum \limits_{k=\Theta_3}^{\Theta_1} \ket{k} \bra{k},\quad\hat{E}_1 = \sum \limits_{k=\Theta_2}^{ \Theta_4} \ket{k} \bra{k},\quad\hat{E}_{\mbox{\scriptsize{fail}}} = \hat{\mathds{1}} - \hat{E}_0 - \hat{E}_1,
\end{equation}
where $\hat{E}_0$ and $\hat{E}_1$ correspond to the outcomes which Bob interprets as `0' and `1', respectively; the operator $\hat{E}_{\mbox{\scriptsize{fail}}}$ corresponds to the failed outcome (the corresponding bit positions will be removed at the post-selection stage); $\ket{k}$ stands for a Fock state containing $k$ photons; $\hat{\mathds{1}}$ is the unity operator, and $\Theta_{1-4}$ are the post-selection parameters that Bob chooses in correspondence to the amount of the observed leak $r_{\mbox{\scriptsize{E}}}$ created by the eavesdropper.

\begin{figure}[h!]
    \centering
     {\includegraphics[width = 0.6\linewidth]{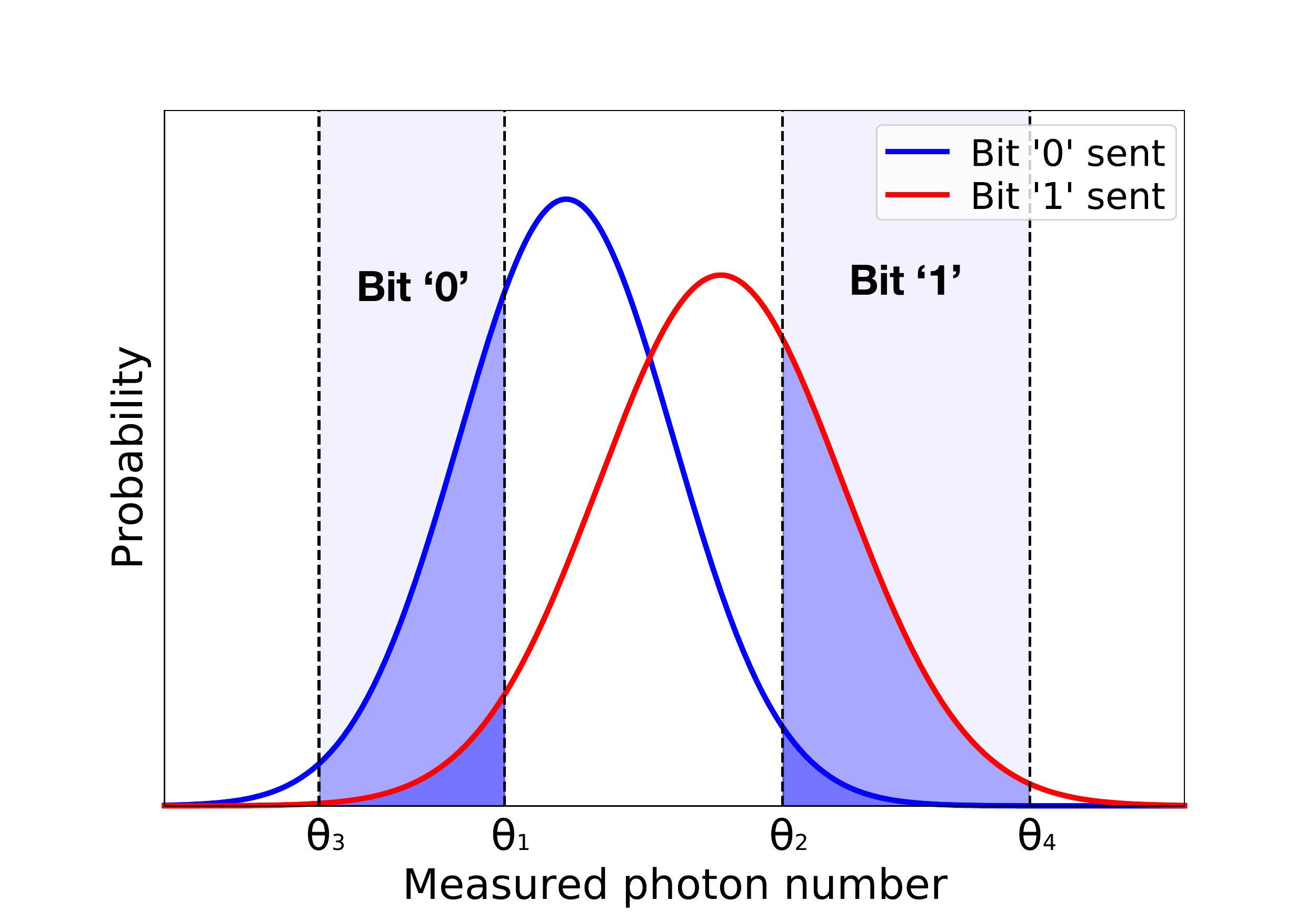}}
    \caption{\textbf{Post-selection.} The sketch shows the probability distributions for the photon numbers corresponding to the sent `0' and `1' bits. The values of photon numbers falling between $\Theta_3$ and $\Theta_1$ are interpreted as corresponding to bit `0', accordingly, those that fall between $\Theta_2$ and $\Theta_4$ constitute the bit `1'. The rest values are discarded as inconclusive results. The interval between $\Theta_1$ and $\Theta_2$ corresponds to the overlap of the distributions implying the noticeable probability of making an error when discriminating between the signals corresponding to different bits. By increasing this interval, we suppress the average error but increase the number of bits to be disregarded. The values that are less than $\Theta_3$ and larger than $\Theta_4$ are the values that eliminate Alice's and Bob's advantage over the eavesdropper, Eve, since optical amplification generates correlations between pulses obtained by Bob and Eve. As a result, the extreme photon numbers at Bob's side appear to be quite distinguishable for Eve as well. Optimizing parameters $\Theta_{1-4}$ enables maximizing the key generation rate and achieves the balance between the `0' and `1' numbers in the sifted sequence.}
\label{postselection}
\end{figure}

The exemplary probability distributions of the measured photon number corresponding to bits `0' and `1' are presented in Fig.\,\ref{postselection}.
Next, Bob sifts the measurement results through four $\Theta_{1-4}$ borders and obtains the raw key.

To correct mismatches between Bob's raw key and Alice's sequence, the error correction procedure is used (see SI, Note\,6 for details).
Next, during privacy amplification legitimate users compress the distributed sequence to eliminate Eve's information about the resulting key.
The compression coefficient is determined by a particular error correction method and by Eve's information obtained from intercepted parts of the bit-encoding states $I(\mbox{A}, \mbox{E})$ (described in the Subsection\,\ref{Eves_info_section}).
The last one depends on the loss control precision.
The length of the resulting key is
\begin{equation}
    L_{\mbox{\scriptsize{f}}}=p_{\checkmark} L\cdot\big(S(\mbox{A})-f S(\mbox{A}|\mbox{B})-I(\mbox{A},\mbox{E})\big).
    \label{final_length}
\end{equation}
Here, $S(\mbox{A})$ is Alice's system entropy provided that the post-selection is carried out; $S(\mbox{A}|\mbox{B})$ is Alice's system entropy obtained under the condition that the results of Bob's measurements are known and that the inconclusive outcomes are discarded.
The quantity $f$ is determined by the error correction code, and $f S(\mbox{A}|\mbox{B})$ is the fraction of the sequence leaked to the eavesdropper during the error correction procedure.
The probability of a conclusive outcome at Bob's is denoted by $p_{\checkmark}$, and $p_{\checkmark} L$ is the length of the sifted bit sequence.
The final secret key must meet the requirement that $L_{\mbox{\scriptsize{f}}} > 0$, otherwise, the distributed sequence is not considered safe and is not added to the general key.

\subsection{Possible attacks and ways of protection}\label{eavesdropper}
Table\,\ref{atacks} presents the possible attacks, their description, and the corresponding ways of protection.

\begin{table}[h!]
    \setlength{\tabcolsep}{4pt}
    \raggedright
    \centering
\begin{tabularx}{\linewidth}{|X|X|X|}
    \hline
    The attack & Brief description & The way of protection
    \\ \hline
    Measurement of the natural losses (see Subsection\,\ref{atack_demon}).
    &
    Rayleigh scattering can cause a leak of the signal. Then the eavesdropper can eavesdrop even without intercepting the line.
    &
    We work with moderate signal power. Corresponding losses constitute a few photons per 100 meters\,\cite{new_theory}, which does not allow measuring them using any realistic photodetectors.
    \\ \hline
    Measuring losses at the splices (see Subsection\,\ref{attack_on_welding}).
    &
    The eavesdropper can collect the signal lighting up out of the cable at the fusion spliced joints without creating additional losses. 
    & We make careful specially optimized splicing procedures. Besides, we demonstrate that not all the losses in the channel go outside the cable.
    \\ \hline
    Creating a local leak (see Subsection\,\ref{atack_attenuation}). 
    &
    The eavesdropper can achieve diverting the part of the signal by e.g. local bending of the fiber channel or splicing an optical beam splitter. 
    &
    We monitor the optical losses in the fiber to detect the eavesdropper.
    \\ \hline
    Creation of the local leak and the increase of the amplifying coefficient in the magistral line amplifier (see Subsection\,\ref{atack_amplifier}).
    &
    The eavesdropper can divert a part of the transmitting signal and compensate for the visible losses by increasing the amplifier power. 
    &
    We design the TQ amplifier settled in an active fiber channel in a way that it always performs in a regime of maximal pumping power and population inversion.
    \\ \hline
\end{tabularx}
    \caption{Possible attacks against the protocol and protection methods.}
\label{atacks}
\end{table}

\subsubsection{Attacks with the measurements of the natural losses}\label{atack_demon}
During the signal transmission along the fiber optical channels, part of the signal scatters out due to Rayleigh scattering. 
Then the eavesdropper can arrange an anomalously large distributed detector next to one of the magistral amplifiers and watch the complete signal even without the mechanical intervention into a transmitting channel. This problem has been described in\,\cite{new_theory}. To avoid this problem we work at the signal intensity that is much less than some minimal critical value at which the eavesdropping is still possible.

\subsubsection{Measuring losses at the splices}\label{attack_on_welding}
During the transmission, a portion of the signal may leak outside the channel at the splice joints. 
The eavesdropper, thus, gets an opportunity of gathering this locally leaking radiation, avoiding, thereby, the creation of any additional losses through mechanical interventions.
However, as shown in the SI, Note\,2, not all the losses go outside the fiber optic communication line. A significant portion of the losses remains confined within the cable and is inaccessible to potential eavesdroppers.
Our estimates suggest that the loss at a splice joint amounts to a mere 0.1\%, which is acceptable for transmissions spanning over 1000\,km.
Notably, one cannot avoid splice joints near amplifiers, but we implement additional protection at these points.

\subsubsection{The attack with creation of the local leak}\label{atack_attenuation}
Using a beamsplitter the eavesdropper can divert and use the part of the signal.
Our fiber optic communication system design uses stringent physical loss control over the communication line which makes it impossible for eavesdroppers to introduce a beamsplitter into the line remaining unnoticed.
The eavesdropper may try to arrange the optical fiber bending and collect all the outcoming from the channel signal. Yet we can detect even the fast eavesdropper's intervention using the regular transmittometry via detecting the sharp integral change of the intensity in the whole line, see Subsection\,\ref{lock_in}.
In addition, the conventional Photon Number Splitting attack can be easily detected, since it requires direct access to the signal for manipulating the pulses which results in change of the channel's tomogram.

\subsubsection{The attack against the amplifier}\label{atack_amplifier}
Diverting the part of the signal using the beamsplitter, the eavesdropper decreases the integral intensity in the line. Therefore, to remain unnoticed the eavesdropper should recover the observable losses using an amplifier.
The eavesdropper cannot introduce its own amplifier since the communication line is permanently subject to loss control and any mechanical intervention will be immediately detected. The retuning amplifying coefficient of the magistral amplifier is also impossible since it works in the regime of the maximal pumping power. Moreover, because of the amplifier security control mechanism any mechanical action immediately increases the losses.

\subsection{Eavesdropper's information estimation}\label{Eves_info_section}
In the context of the proposed loss control approach, the eavesdropper is not able to conduct conventional attacks including the replacement of the optical line with the ideal (lossless) quantum channel, since such attacks dramatically change the line reflectogram and are to be easily detected.
Thus, Eve has to introduce local losses at some point in the line.

In this work, we analyze in detail the situation in which Eve's intrusion point is located right after Alice's lab.
In such a case, Eve's quantum states corresponding to different bit values do not depend on Bob's choice of post-selection procedure. 
Let $r_{\mbox{\scriptsize{E}}}$ be the minimal detectable artificial leakage and $\gamma_a$ be the amplitude of the bit-encoding pulse.
Eve's density matrix is
\begin{equation}
    \hat{\rho}_{\mbox{\scriptsize{E}}}^{(a)}
    =
    \frac{1}{2\pi}\int\limits_{0}^{2\pi}d\varphi\,\,
    \left|e^{i\varphi}\sqrt{r_{\mbox{\scriptsize{E}}}}|\gamma_a|\right\rangle \left\langle e^{i\varphi} \sqrt{r_{\mbox{\scriptsize{E}}}}|\gamma_a| \right|_{\mbox{\scriptsize{E}}}
    =
    e^{-r_{\mbox{\tiny{E}}}|\gamma_a|^2}
    \sum\limits_{n=0}^{+\infty}\frac{\left(r_{\mbox{\scriptsize{E}}}|\gamma_a|^2\right)^n}{n!}\ket{n}\!\bra{n}_{\mbox{\scriptsize{E}}}.
    \label{eve_matr_ints}
\end{equation}
The integration over all possible values of phase factor corresponds to the phase randomization procedure applied by Alice.
If the eavesdropper intervenes at an arbitrary point of the optical line, the amplifiers located before the point induce correlations between Eve and Bob. 
The detailed analysis of the correlations' influence within an idealized model is provided in Ref.\,\cite{new_theory}, while an analysis of laser instability and noise optical modes as sources of additional correlations and noises will be the subject of subsequent publications.

The maximum amount of information that Eve can extract from the obtained states\,(Eq.\,(\ref{eve_matr_ints})) on average is upper bounded by the Holevo quantity\,\cite{Holevo} $\chi$ which, in our case, can be expressed as
\begin{equation}
    I(\mbox{A},\mbox{E})\leq\chi
    =
    S\left(\frac{p\left(\checkmark|0\right)}{2p_{\checkmark}}\hat{\rho}_{\mbox{\scriptsize{E}}}^{(0)}+\frac{p\left(\checkmark|1\right)}{2p_{\checkmark}}\hat{\rho}_{\mbox{\scriptsize{E}}}^{(1)}\right)
    -
    \frac{p\left(\checkmark|0\right)}{2p_{\checkmark}}S\left(\hat{\rho}_{\mbox{\scriptsize{E}}}^{(0)}\right)
    -
    \frac{p\left(\checkmark|1\right)}{2p_{\checkmark}}S\left(\hat{\rho}_{\mbox{\scriptsize{E}}}^{(1)}\right),
    \label{eve_holevo}
\end{equation}
where $S(\hat{\rho})=-\mbox{tr}\left[ \hat{\rho} \log_2 \hat{\rho} \right]$ is von Neumann entropy, $p(\checkmark|a)$ is the probability of a conclusive measurement result at Bob's end provided that the sent bit is $a$ and the sum $p_{\checkmark}=p\left(\checkmark|0\right)+p\left(\checkmark|1\right)$ determines the average probability of a conclusive result.
The value $I(\mbox{A},\mbox{E})$ includes information that Eve extracts from the measurement of the intercepted photons and does not comprise information leaked during the error correction procedure.
Notably, if the legitimate users are able to construct an upper bound on the potential eavesdropper's quantum memory coherence time, the estimation can be improved by the means of encrypting classical information disclosed during error correction procedure\,\cite{encrypting_error_correction}.

\bigskip
\subsection{Experimental key generation rate}\label{generation_speed}

In the key distribution process, the transmittometry pulses are generated in consecutively with the bit message. The repetitive key distribution and transmittometry steps are interrupted by a reflectometry step. The reflectometry process takes half the time of the key distribution in the line. Thus, the speed of sending the raw uncorrected and not post-selected key is 200\,kbps.

After Bob receives the signal using the ADC, further processing of the key consists of sequence post-selection and error correction. About 99\% of the sequence is discarded in the post-selection process. In the error correction process, 80\% of the post-selected sequence is discarded. Finally, it is also necessary to take into account the fraction of the signal that Eve receives if it takes a part of the optical signal less than the accuracy of the line loss control. With our setup parameters, with a bit '0' intensity of 12,300 photons and a bit '1' intensity of 13,700 photons, Eve's information is no more than 10\% of the post-selected key (see Subsection\,\ref{Eves_info_section}). Thus, the final generation rate is 168 bps.

\subsection{The encountered problems}
Table\,\ref{problems} summarizes the difficulties and problems that our realized protocol has overcome. 

\begin{table}[!h]
    \centering
    \begin{tabularx}{\linewidth}{|X|X|X|}
    \hline
    The problem & Brief description & Solution    
    \\ \hline
    An exponential decay of the signal in an optical fiber (see Subsection\,\ref{Low_intensity_defence}).
    &
    The signal exponential decay in the optical channel results in the long-distance communication problem.
    &
    The use of the EDFA-like amplifiers was modified to fit our protocol.
    \\ \hline
    Detecting low-power information-carrying signal\, (see Subsection\,\ref{problem:low_intensity}).
    &
    Our QKD protocol design implies low-power key distribution.
    &
    The TQ-made amplifier having an amplifying coefficient of 20\,dB and a low signal-to-noise ratio is developed.
    \\ \hline
    The high level of the ASE from the EDFA (see Subsection\,\ref{problem:high_ASE}).
    &
    The EDFA randomly generates wide-spectrum optical irradiation; long optical lines contain many EDFAs which˙results in a high noise level.
    &
    The TQ-made narrow-band, 8.5 GHz, thermostabilized filter is developed.
    \\ \hline
    Laser irradiation generation in a long optical line without optical isolators (see Subsection\,\ref{problem:generation}).
    &
    In long optical lines having active fiber segments in TQ-EDFAs, the unbalanced loss and amplification levels may cause laser irradiation due to multiple reflections at connectors and Rayleigh scattering. 
    &
    The developed optical line has high control over the loss and amplification levels upon adding every amplifier, having enabled the right choice of the amplification coefficient corresponding to losses.
    \\ \hline
    Floating of the offset voltage at the amplitude modulator (see Subsection\,\ref{problem:AM_drift}). 
    &
    The working point of the voltage offset, i.e., the working point of the amplitude modulator may shift with time.
    &
    The part of the optical irradiation is diverted and controlled using an additional detector at Alice's side. Then the voltage is tuned according to the detector's indications.
    \\ \hline
    \end{tabularx}
\caption{The encountered problems and the ways of their solution.}
\label{problems}
\end{table}

\subsubsection{Exponential signal decay in the optical fiber}\label{Low_intensity_defence}
Upon spreading along the optical fiber, the optical signal exponentially decays with the distance. Therefore, for long-distance communication, the TQ-EDFAs are used. The TQ-EDFA ensures the loss compensation at a distance of about 50\,km. Since this amplifier is not an ideal quantum repeater, it generates additional noise. Importantly, the TQ-EDFA design is adjusted to our protocol, see the details in SI, Note\,1.

\subsubsection{Detecting the low power level}\label{problem:low_intensity}
Since, as has been mentioned above, a high power level excludes the possibility of the secret information, see Subsection\,\ref{atack_demon}, the information transmission is executed at a moderate intensity of the optical signal. Hence this signal is to be amplified before sending it to the high-sensitivity detector. To that end, our team developed a specific preamplifier with an amplifying coefficient of 20\,dB. An additional constructive feature of the TQ amplifier that differs it from the magistral amplifiers is the presence of optical isolators in its constructive design.

\subsubsection{The high ASE level}\label{problem:high_ASE}
A long-distance optical communication line exploits magistral amplifiers EDFA to compensate for the exponential optical losses. Since EDFAs are not the ideal quantum repeaters, they generate ASE leading to the noise interfering with the detected information-carrying signal. We have developed a narrow-band 8.5\,GHz filter that has enabled to decrease the ASE power arriving at Bob's detector. An important technological component of the new TQ narrow band filter is a thermal stabilizer that has enabled the $\pm 0.01 \mbox{ K}$ of thermal stability precision, also developed by our team.

\subsubsection{Laser irradiation generated in the long-distance line not containing optical isolators}\label{problem:generation}
\begin{figure}[h!]
\centering
    {\includegraphics[width = 0.6\linewidth]{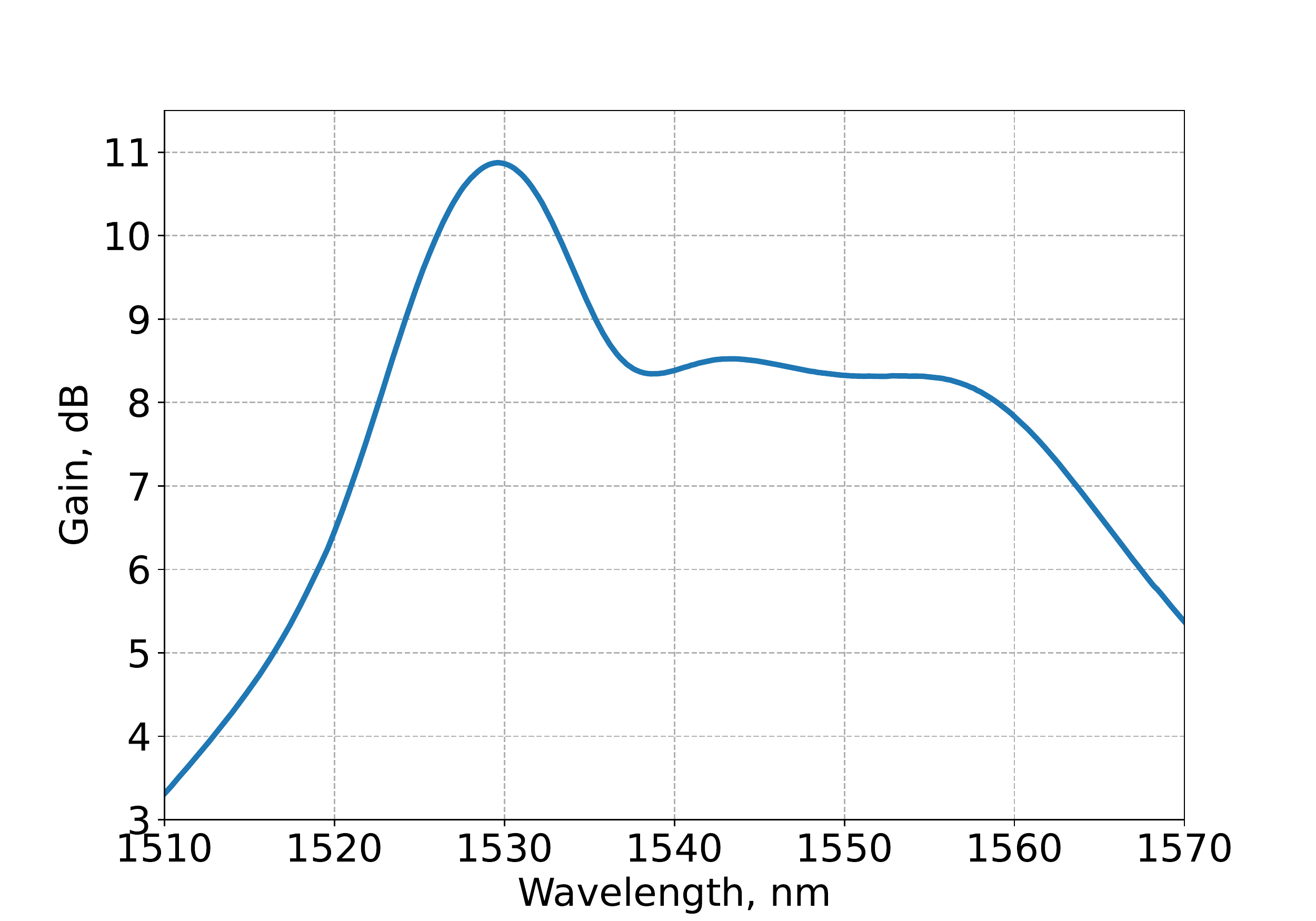}}
    \caption{\textbf{Amplification spectrum of magistral amplifiers}. The amplification peak for the amplification coefficient of about 10\,dB is at 1530.33\,nm.}
\label{amplif_spectrum}
\end{figure}

A long-distance optical channel exploits a significant number of EDFAs and an amplification effect is hosted by the erbium-doped fiber segments; these segments are referred to as active medium regions. Since in the optical line, several signal reflections occur the whole line is to be viewed as an active medium containing weak resonators. Therefore, the transfer of the ASE critical power is accompanied by the generation of the irradiation analogous to laser irradiation. We eliminate this parasitic generation by choosing the regime of the information signal transmission ensuring being in the ASE amplification peak, see Fig.\,\ref{amplif_spectrum} and using the TQ-EDFAs-designed to ensure the amplification coefficient of 10\,dB providing thus losses compensation over the 50\,km distance.

\subsubsection{Floating of the offset voltage at the amplitude modulator}\label{problem:AM_drift}
The possibility of the offset voltage floating resulting in the shift of the working point (often referred to as a bias point) of the amplitude modulator, leads also to the change of the average intensity of the optical signal. To compensate for the bias point shift, our TQ team has developed an algorithm that uses the data from the monitoring detector, shown in Fig.\,\ref{QKDscheme}, and returns the changing average intensity of the optical signal to its original magnitude.

\bigskip\section{Line control}\label{control_section}
The important element of the protocol is the control over the losses in the communication line enabling the legitimate users to momentarily estimate the information stolen by eavesdroppers.
The loss control method is universal and can be applied to a wide range of QKD protocols boosting their performance\,\cite{boosting}.

\subsection{Transmittometry}\label{lock_in}
\begin{figure}[h!]
    \centering
     {\includegraphics[width = 1.0\linewidth]{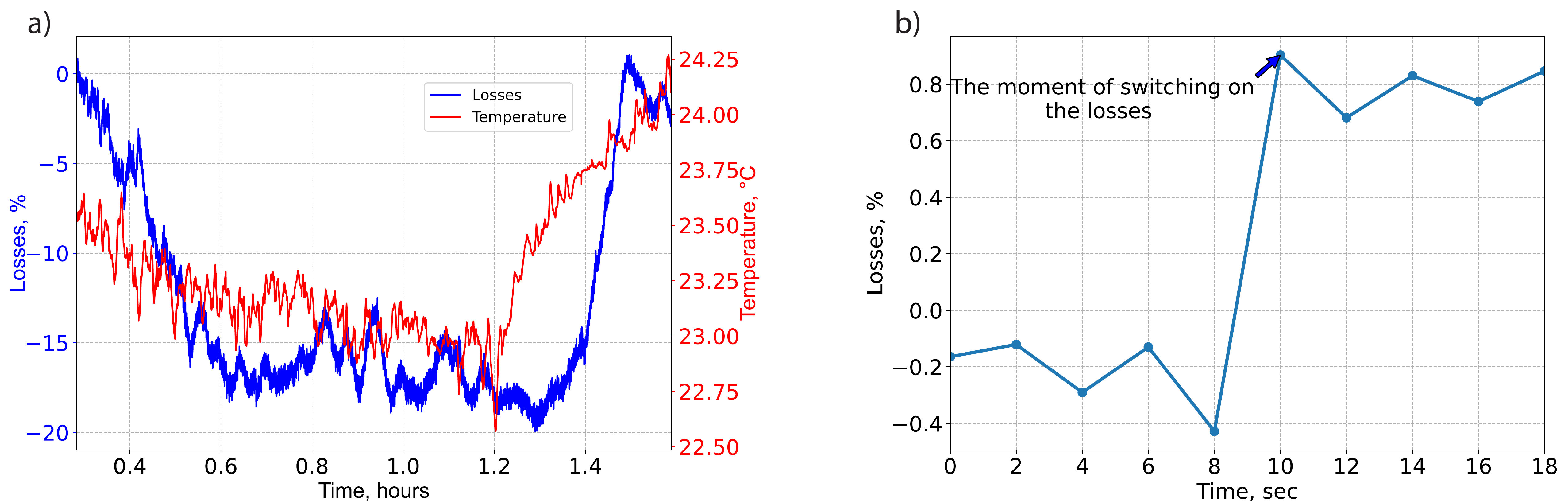}}
    \caption{
    \textbf{(a)} 
    Time dependence of the loss in the communication line and room temperature. Correlations between the temperature change and line loss can be found.
    \textbf{(b)} 
    Measurements detecting the intervention into the communication line. The associated introduced losses are about 1\%.}
    \label{lockin_stabil_loss}
\end{figure}

The control over the throughput losses is executed as follows. The sinusoidal-modulated signal with the frequency of 25\, MHz is sent along the optical line. The pulse length of every control sending is 1\,ms. At the line output after the TQ-manufactured preamplifier and optical filters, the signal is received by the detector.
This approach is analogous to the lock-in method\,\cite{lock_in} where the signal taken from the detector is received by an oscilloscope and transferred to the user's computer.  Further treatment is the search, at the modulation frequency, for the amplitude of the signal Fourier transforms at the reference frequency. The ratio of the obtained and reference amplitudes enables us to determine the magnitude of the losses. 
The reference loss magnitude is updated at each reflectometry session, so we get the amplitude of the sine wave $a_{\mbox{\scriptsize{ref}}}$ passing through the optical line. Further, at each iteration of transmittometry, after finding the amplitude of the signal $a_{\mbox{\scriptsize{t}}}$ at the desired frequency, we can obtain the loss value as $1 - a_{\mbox{\scriptsize{t}}}/a_{\mbox{\scriptsize{ref}}}$. Figure\,\ref{lockin_stabil_loss}(a) shows the time dependence of loss and room temperature. The temperature dependence of the loss in an individual spool can be found in the SI, Note\,4. For the obtained data, the standard deviation is 0.2\%, provided that we perform reflectometry and update the amplitude reference value every few iterations of transmittometry. Therefore, this is the limit of the accuracy of Eve's signal measurement.

To model the eavesdropper presence, one introduces the losses into the line (in the implemented measurements the losses are 1\%) and observes the loss coefficient time dependence. The moment of switching on the losses is marked by the arrow, see Fig.\,\ref{lockin_stabil_loss}(b).

The loss control does not necessarily have to utilize the dedicated test pulses. 
Instead, it can be based on analysing the intensities of the information-carrying pulses, making this procedure more similar in its spirit to the conventional QKD.
We plan to pursue this approach in the near future.

\begin{figure}[h!]
    \centering
    \includegraphics[width = 1.0\linewidth]{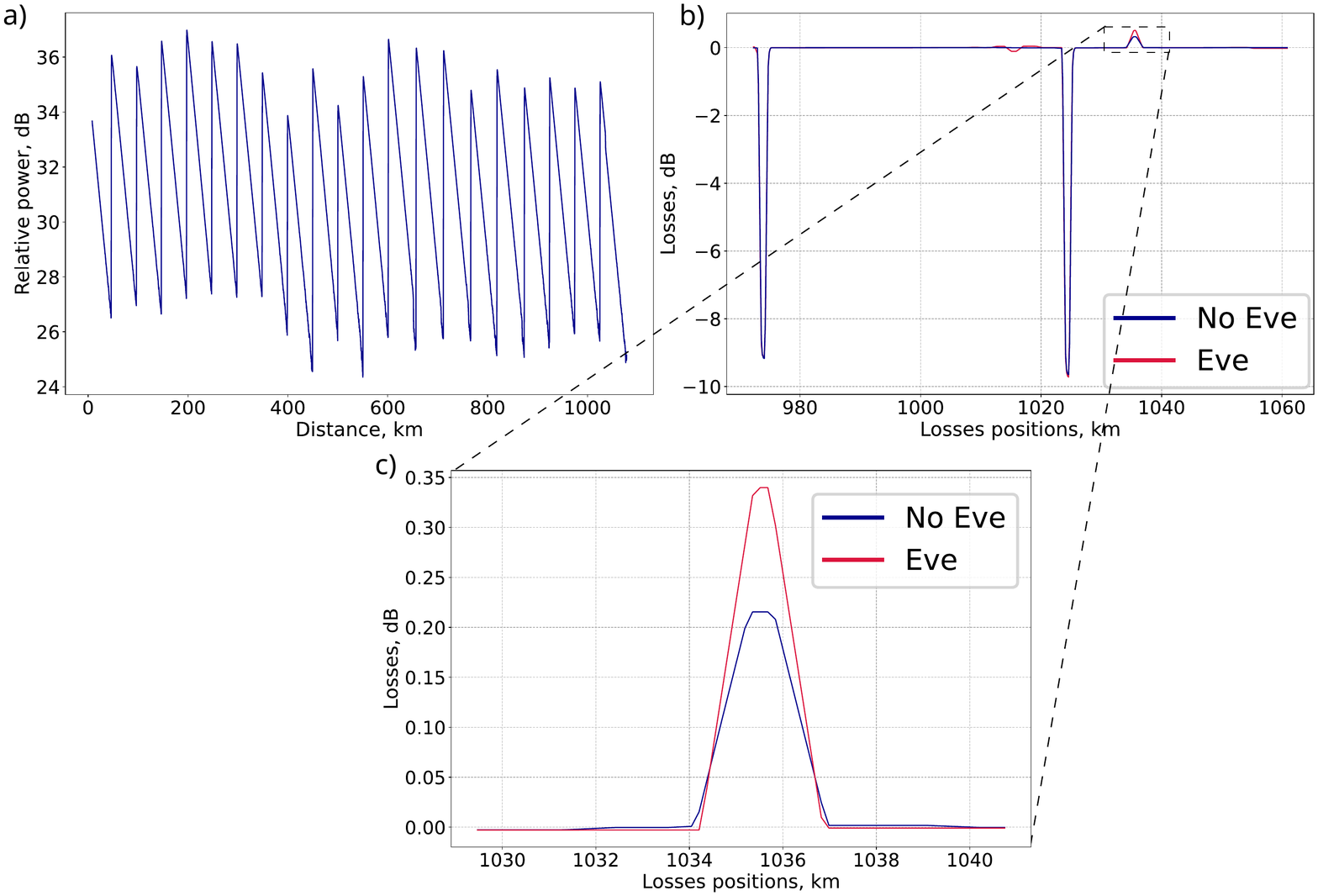}
    \caption{{\textbf{Leakage detection experiment. (a)} The full reflectogram of the communication line (without introduced losses) comprising the 50\,km fiber spools, the EDFAs, and the additional 10\,km spool after the 20th amplifier. The additional spool was set to make an exposed point for eavesdropping. The measurements were carried out with a probing pulse duration of 3\,$\mu$s and averaging over about $2^{12}$ pulses which corresponds to a total measurement time of approximately 2\,mins. The EDFA-induced up-jump of the signal power is clearly seen at every $\sim$\,50\,km. The reflectogram reveals the fusion splice on a 1040 km position. \textbf{(b)} Marked losses. Each pike corresponds to drop in power at this point. The plot is obtained by applying L1-filtering technique and taking discrete derivative. At the 975\,km and 1025\,km distance from the reflectometer, one can notice the "negative losses" due to the EDFA. At the distance of 1035\,km from the reflectometer one can see the initial losses and the additionally introduced $\sim$\,3\% losses caused by the bending of the fiber. \textbf{(c)} The magnified segment where additional losses are introduced. One sees the growth of the losses near the 1035\,km.}}
    \label{refl_expr_res}
\end{figure}

\subsection{Reflectometry}\label{REFL_section}
The OTDR approach can be used for detecting local interventions in the communication line. The procedure consists of sending a short powerful optical pulse into the line and measuring the backscattered signal. The local losses usually occur in optical connections, splices, or bends, which are seen as the main points where the signal leaks out of the line. Although splices only scatter some of the loss outward, see SI, Note\,2, and our line is designed to contain no optical connections in order to prevent excessive leakage, there are still unavoidable external losses that must be estimated. Rayleigh scattering, which is almost impossible for the eavesdropper to exploit, helps to detect faults along the line. If the eavesdropper diverts the part of the signal, the intensity of the backscattered radiation drops in the segment containing the intervention point.
This allows local interception to be detected.

We present a summary of the current status of OTDR monitoring of the entire line.
Using a special TQ-made OTDR device with 1530 nm laser source and 1280 km distance range, we performed a monitoring procedure for our test line. {Our device is compatible with bidirectional optical amplifiers.} Figure\,\ref{refl_expr_res} represents the results of the measurements. 
During the experiment, external losses were detected in the optical line at almost maximum range. The local leakage was detected and evaluated at 3\%. 

The collected data enable us to conclude that the power of the detected signal due to the Rayleigh scattering does not decay upon going through amplifiers along the whole communication line. This follows from the fact that the peaks presented in Fig.\,\ref{refl_expr_res} do not lose their height on average. 
This result allows us to justify using an OTDR approach and further advance this technique for our system.

\begin{figure}[h!]
    \centering
    \includegraphics[width=1.0\linewidth]{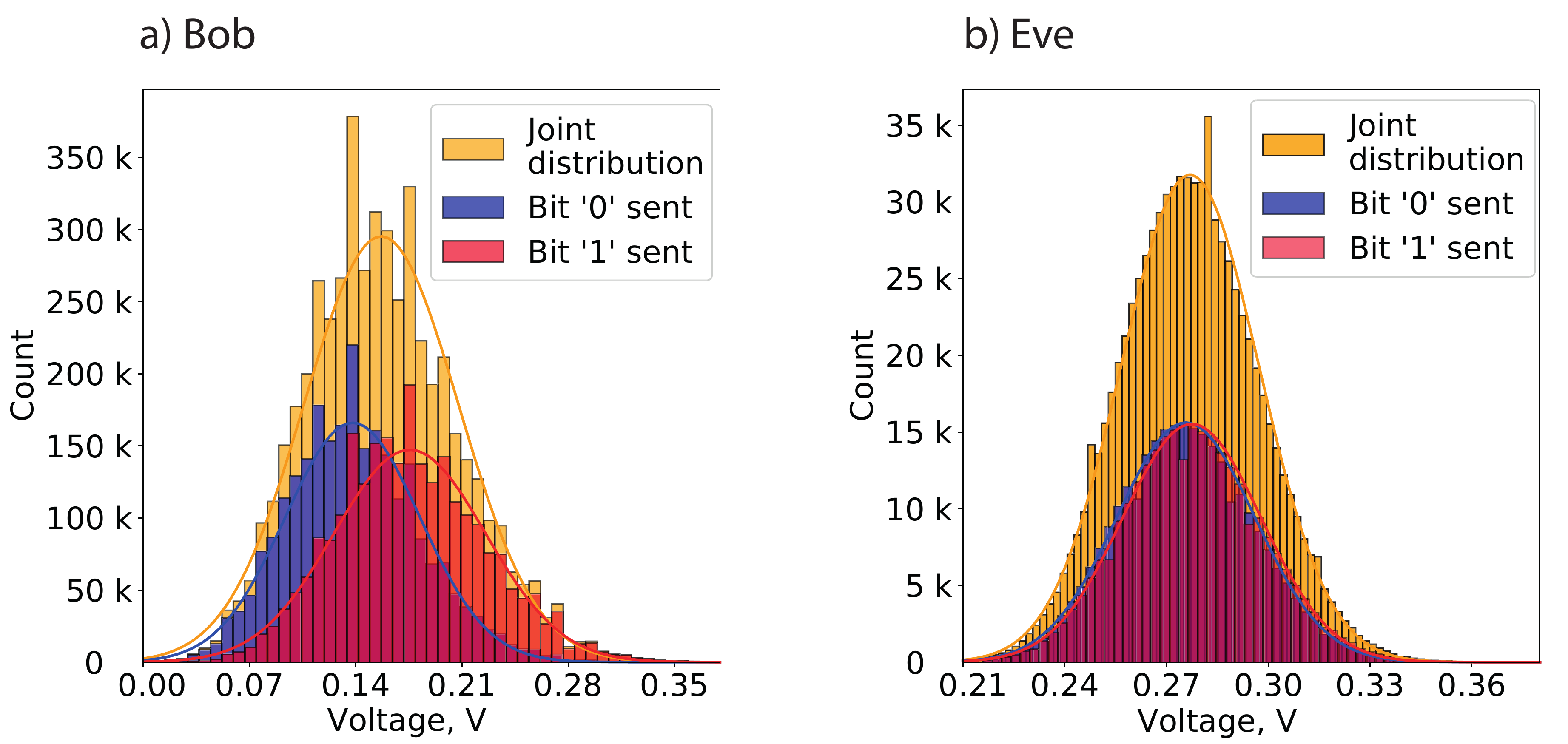}
    \caption{
    \textbf{Voltage distributions conditioned on bits `0' and `1' for Bob and Eve at the end of the line.}
    Bob and Eve measure the intensity of the received signal, which corresponds to the number of photons in the current state, and transform it into the voltage. Note that Bob's TQ-EDFA gain is 170 times less than Eve's.
    The distribution of the signals corresponding to Alice's `0' is shown in blue, and Alice's `1' is shown in red. Yellow represents the total distribution of the signal intensities. Each distribution is approximated by normal Gaussian distributions of the corresponding colors. Note some deviations from the symmetric Gaussian curves. \textbf{(a),} The signal distribution for Bob. 
    The approximate parameters of the shown distributions are as follows: the peaks for `0', `1', and `sum' are seen at the final stage of detection on the oscilloscope, $V_0 = 0.138\mbox{ V}$, $V_1 = 0.176\mbox{ V}$ and $V_{\mbox{\scriptsize{sum}}} = 0.157\mbox{ V}$, respectively, while the standard deviations are, respectively, $\sigma_0 = 0.044\mbox{ V}$, $\sigma_1 = 0.050\mbox{ V}$ and $\sigma_{\mbox{\scriptsize{sum}}} = 0.051\mbox{ V}$. \textbf{(b),} Illustration of the distribution of `0' and `1' for Eve.
    Originally, the number of photons in the bits sent by Alice was 11,400 photons for bit `0' and 15,100 photons for bit `1'.
    }
\label{distribution_1}
\end{figure}

\bigskip\section{Illustration of advantage over the eavesdropper}
As shown above, the current version of OTDR allows to detect the local interventions at distances more than 1000\,km.
Here we illustrate the advantage of Bob, who measures the intensity of the received signals, over Eve being located at the end of the line and attempting to distinguish between intercepted parts of bit-encoding states.
In this experiment, simulated eavesdropper has an access to 1\% of the transmitted optical pulse.
Since the electromagnetic signal, as discovered by Max Planck, is quantized and carried by discrete particles, photons, the photon number in the pulses corresponding to `0' and `1' states intercepted by Eve is 100 times less than Bob's corresponding photon numbers. Notably, for the quantum coherent states that we use for encoding `0' and `1' bits there are inevitable quantum fluctuations in the measured pulse intensities. Therefore, the less the number of photons $N$ that Eve manages to intercept, the bigger the relative fluctuations which scale as $1/\sqrt{N}$ are. To visualize this effect,
we plot the experimentally obtained distributions of the `0' and `1' states as functions of voltage corresponding to the intensity of an incoming signal, using the statistics obtained for a large number of the secret quantum keys, 
at Bob's side, Fig.\,\ref{distribution_1}(a). 
For Eve we measure the statistics of the number of photons received during the attack using our equipment, Fig.\,\ref{distribution_1}(b). 
The measurement is made for the signal coming to Bob, but attenuated by a factor of 100. 
In order to distinguish the signal coming to Eve, the amplification factor was increased.
In this experiment, Alice encodes the bit `0' via the state containing 11,400 photons and the bit `1' via the state containing 15,100 photons.
One sees that while at Bob's side the `0' and `1' states are still distinguishable, at Eve's side two corresponding distribution functions completely overlap making the determining of the bit state impossible. Bob knows only the total signal distribution, where he has to put in the post-selection parameters (boundaries) to ensure efficient discrimination between the states. The distance between the distribution peaks depends on the difference between the intensity of the signals. 

Thus, Bob's informational advantage is achieved precisely because of the quantum mechanical fluctuation limit for the coherent state. For this reason, the eavesdropper has no way of gaining informational advantage by measuring a small fraction of the original optical signal. At the same time, as we noted earlier, the eavesdropper cannot obtain any significant part of the optical signal without making a noticeable for loss control introduction into the line.

We acknowledge that Eve might possess equipment of higher precision than that of users. 
The degree of distinguishability of the states’ components seized by Eve is reflected in their scalar product: 
\begin{equation}\label{Overlap}
    \left\langle\sqrt{r_\text{E}}\gamma_0|\sqrt{r_\text{E}}\gamma_1\right\rangle=\exp \left(-r_\text{E}(|\gamma_0|-|\gamma_1|)^2\right)
\end{equation}
(as shown in Ref.\,[\citeonline{new_theory}], due to phase randomization and amplifiers' action the states become mixed, but here we ignore this fact for ease of explanation).
Taking the parameters of the pulses as utilized in the current implementation, $|\gamma_0|^2=12,300$, $|\gamma_1|^2=13,700$, and leakage $r_\text{E}=0.6\%$ (equivalent to three variances in our experiment), the overlap of Eve's states is $0.8$.
Such overlap ensures low distinguishability, causing high error rate at Eve's end.
This underscores the quantum nature of the realized protocol.

The numbers $|\gamma_0|^2$ and $|\gamma_1|^2$ are determined from the pulses' energies $E_0$ and $E_1$:
\begin{equation}\label{h_formula}
    |\gamma_{0(1)}|^2 \simeq \frac{E_{0(1)}}{h\nu},
\end{equation}
where $\nu$ is the frequency of optical radiation, and $h$ is Planck's constant.
This relationship provides yet another lens on the quantumness of our protocol: the indistinguishability of quantum states lies in the inherent discrete property of optical radiation and the limited photon count within the signal.
In a classical limit where Planck's constant $h$ tends to zero, $h \rightarrow 0$, the overlap given by Eq.\,(\ref{Overlap}) becomes zero for any states' energies except for the trivial case where $E_0=E_1$.
Yet, with a finite value of $h$, quantum mechanics manifest in non-orthogonality of the information carrying states, as described above in this paper. 

\bigskip\section{Key randomness analysis}\label{bitber_section}

To check to which extent our final key corresponds to the binomial distribution, we compared the corresponding values of the mean and the standard deviation (see Table\,\ref{tab:binom_compare}). Here, we use the selection of the 8000 bit sequences of size 7500 bits.

\begin{table}[h!]
    \centering
    \begin{tabular}{|l|l|l|}
    \hline
    Value & Theoretical  & Experimental \\ \hline
    Mean  & 3750.00  & 3750.55      \\ \hline
    Standart Deviation & 1875.00 & 1848.90      \\ \hline
    \end{tabular}
    \caption{Theoretical and experimental values of mean and standard deviation.}
    \label{tab:binom_compare}
\end{table}

We have also executed the tests verifying the randomness of the distributed key.
The detailed analysis has been performed by NIST tests using official program 'NIST STS', as shown in Table\,\ref{nist}. 

\begin{table}[h!]
    \centering
    \begin{tabular}{|l|l|l|l|}
    \hline
    Statistical Test        & P-value & Proportion & Assessment \\ \hline
    Frequency               & 0.544254        & 0.995           & Success    \\ \hline
    Block Frequency          &  0.342451      & 1.000           & Success    \\ \hline
    Cumulative Sums          &  0.129620      & 0.985           & Success    \\ \hline
    Runs                    & 0.595549        &  0.980          & Success    \\ \hline
    Longest Run              & 0.289667        & 0.985           & Success    \\ \hline
    Rank                    & 0.392456        & 0.985           & Success    \\ \hline
    FFT                     & 0.946308       &  0.995          & Success    \\ \hline
    Non Overlapping Template  & 0.045675        & 0.985           & Success    \\ \hline
    Overlapping Template     & 0.605916        & 0.995           & Success    \\ \hline
    Universal               & 0.605916        &  0.985          & Success    \\ \hline
    Approximate Entropy      & 0.807412       &  0.995          & Success    \\ \hline
    Random Excursions        & 0.082177        & 0.989           & Success    \\ \hline
    Random Excursions Variant & 0.071670        & 0.989           & Success    \\ \hline
    Serial                  & 0.133404        & 0.990           & Success    \\ \hline
    Linear Complexity        & 0.118812        &  0.980          & Success    \\  \hline
    \end{tabular}
    \caption{Result of NIST tests over a 90 MBit final key.}
\label{nist}
\end{table}
As the default significance level $\alpha = 0.01$, so the proportions of sequences that satisfy P-value $> 0.01$ should be greater than 0.96, our final key successfully passed NIST tests. 

\bigskip\section{Conclusion and discussion}
In this study, we presented the first practical implementation of the TQ-QKD protocol as delineated in Ref.\,[\citeonline{new_theory}].
Utilizing our theoretical foundation, we achieved quantum key distribution over an unprecedented distance of 1079 km at a rate of 168 bps, as illustrated in Fig.\,\ref{comparing}. 
This accomplishment has been achieved through the development of the specific optical amplification and line control equipment.
Crucially, we have highlighted potential enhancements for this equipment, anticipating longer transmission spans in forthcoming studies. 
These advancements pave the way for the future scalable QKD networks.

The key feature of the implemented QKD protocol is that the users control the transmission channel, which may appear unconventional.
Nevertheless, given that the eavesdropper must still distinguish between non-orthogonal quantum states, the realized protocol relies on quantum mechanics and belongs to a broader class of the QKD solutions.
Unlike the computational problems which classical cryptography is based on, the task of distinguishing non-orthogonal quantum states cannot be solved or simplified by future technical progress. 
Our solution is quantum also in the sense that it provides everlasting security of the keys distributed\,\cite{Renner_Wolf}.

The TQ-QKD protocol lifts some of the QKD limitations recently emphasized in the debate between the NSA and 
the scientific community\,\cite{Renner_Wolf, Sych_debate}.
The realized protocol, though it is more device-dependent than other conventional QKD schemes, allows to achieve global transmission distances without employing trusted nodes.
We recognize that increased device-dependence introduces additional risks. 
With that, our approach allows us to achieve unparalleled key rates.
Finally, by modifying the model of the key generation process, we bring in an advanced theoretical description of the QKD and its experimental realization.
A comprehensive security analysis, considering the finite-size effect, will be the subject of a dedicated theoretical paper.

\bigskip\section{Data availability}

All of the data that support the findings of this study are available in the main text or Supplementary Information. Source data are available from the corresponding authors on reasonable request.

\section*{Acknowledgements}
We are delighted to thank Dmitry Kronberg and Denis Sych for illuminating discussions.

\section*{Author contributions statement}
M.P., N.K., and V.V. conceptualized the work. 
The experiment was designed by N.K., A.A., and I.Z., while A.A., V.S., and I.Z. assembled the experimental setup and collected data. 
A.S. and M.Y. gathered data with optical time-domain reflectometer. 
The software development was done by I.Z., D.S., and A.O., whereas D.S. took charge of the hardware's design and assembly. 
The underlying theory and interpretation on the experimental data was developed by V.P., A.K, and N.K.
All authors participated in writing the manuscript.
A.B., A.S., and M.Y. wrote the Supplementary Information.

\newpage
\noindent\textbf{\Huge{Supplementary Information}}

\bigskip\section*{NOTE 1. Amplifiers}\label{amplifier_section} 
\begin{figure}[h!]
    \centering
    \includegraphics[width = 0.6\linewidth]{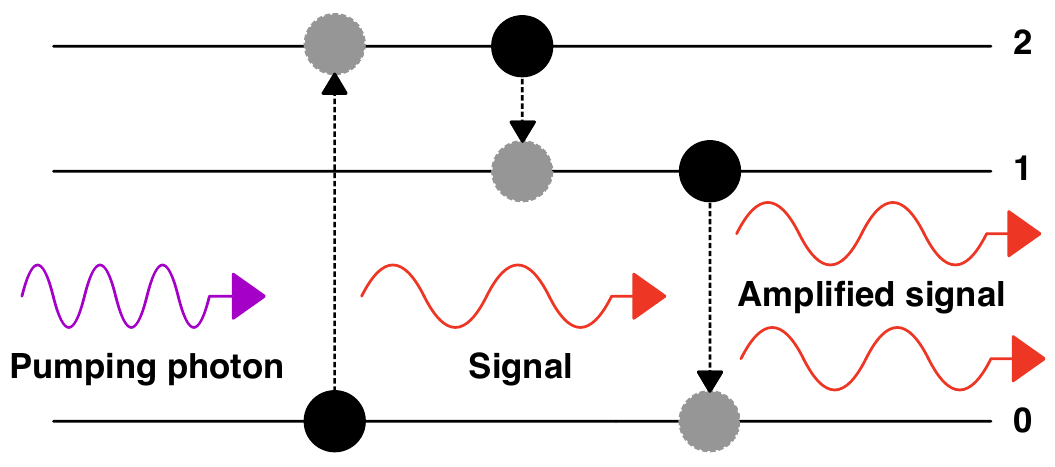}
    \caption{\textbf{Transitions in the erbium-doped fiber system.} Because of the pump absorption, photons from the ground state move to the short-living (about 20 $\mu$s) state. Then photons transit to the first excited state and stay there for about 1\,ms. This way, we keep the inverse population of erbium ions. The transition from state 1 leads to the coherently synchronized photons emission.}
    \label{3dlvl_system}
\end{figure}

Due to natural losses of photons, there is a fundamental limit to long-distance quantum communication. 
The well-known PLOB-repeaterless bound\,\cite{plob} emerges due to the exponential signal decay over distance. It imposes restrictions for creating a wide-spread quantum network. That is the main reason why we should use quantum repeaters which gain incoming signals and compensate for losses in the line.
As an amplifying factor in standard telecommunication routes, the active fiber is used. The most widespread amplifiers are the Erbium-doped fiber amplifiers (EDFA)\,\cite{EDFA_Desurvier} where each erbium ion could be considered as a third-level system depicted in Fig.\,\ref{3dlvl_system}.

\begin{figure}[h!]
\centering
\includegraphics[width = 0.6\linewidth]{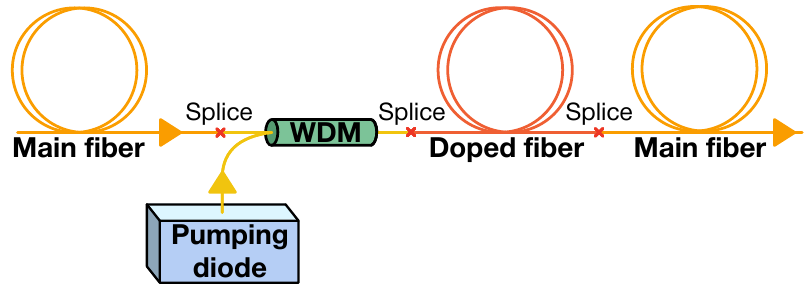}
\caption{\textbf{The scheme of the bidirectional (Terra Quantum AG-made) amplifier.} Wavelength-division multiplexing (WDM) system on its input unites the informational signal from the optical line and pump from the diode and releases them together in an active fiber section which is further connected with the main fiber.}
\label{amplifier_construction}
\end{figure}

We emphasize that our Terra Quantum AG-made optical amplifiers based on the EDFA principle are bidirectional and have no optical isolators, filters, or detectors. Such construction had previously been used in works\,\cite{2000_bidirect_ampl_1, 2000_bidirect_ampl_2, 2000_bidirect_ampl_3} and gives us an additional opportunity to control line via the OTDR. The amplifiers consist of a pumping diode and active fiber which are connected via the wavelength-division multiplexing (WDM) represented in Fig.\,\ref{amplifier_construction}. The system of WDM allows the transmission of signals with different wavelengths onto a single optical fiber at the same time.

Despite we removed isolators from the amplifiers, the line could still stay stable because the amplifiers are spaced far apart and the reflection back does not overcome lasing generation threshold. In the presented 1032-km line we use 19 such amplifiers separated from each other by 50 km. It is important to note here the peak of the EDFA spectrum lies at the wavelength 1530\,nm, therefore, due to removing filters from the amplifiers, we chose 1530\,nm as the main informational wavelength.

Eavesdroppers could try to attack our amplifiers by increasing the gain factor and stealing extra photons. However, the amplifiers work in the regime with almost maximal pumping. It means that we need a minimum number of excited ions to achieve the target amplification of the incoming signal.

\bigskip\section*{NOTE 2. The splice losses}\label{loss_section}
\subsection*{Mode losses caused by splice joint}
\begin{figure}[h!]
    \centering
    \includegraphics[scale=0.08]{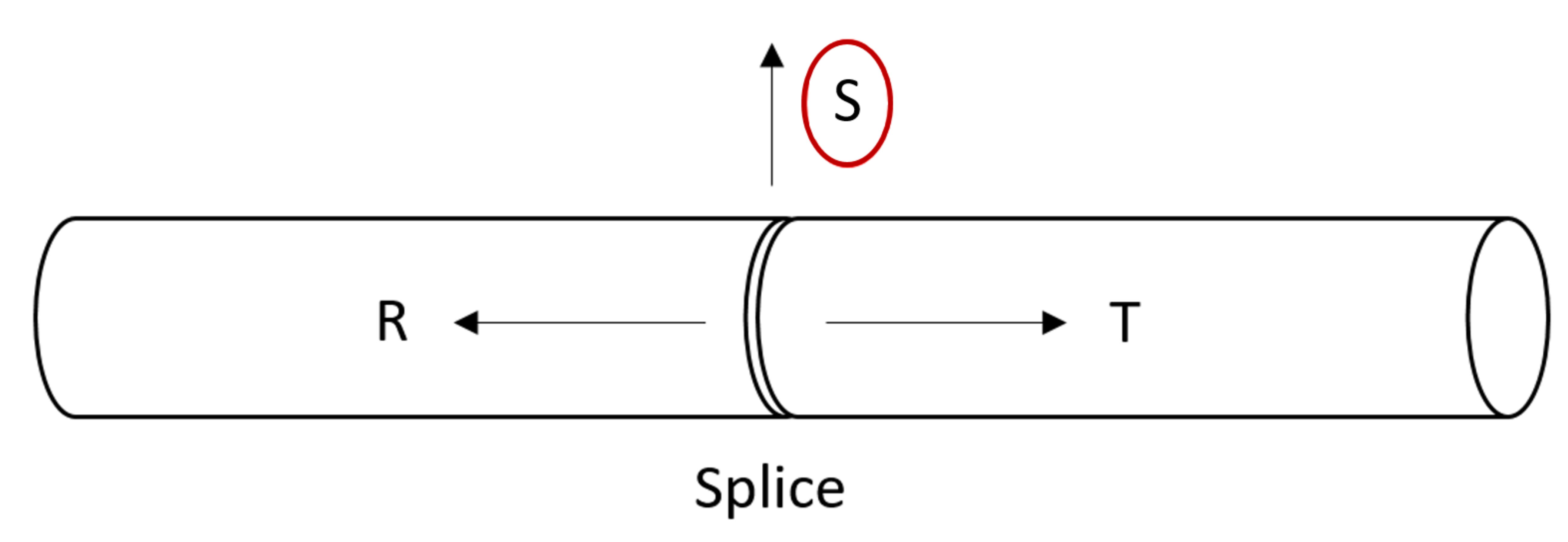}
    \caption{Splice joint in an optical line.}
    \label{general_splice}
\end{figure}

It is usually believed that at present the low-loss splices can be done only in a laboratory, but due to the constantly improving splicing equipment, it becomes useful to measure the loss value for splices done on some commercial device. First of all, we need to consider the part of the optical line with one splice joint (Fig.\,\ref{general_splice}).

It divides this segment of the fiber into two parts. Let $T$, $R$, and $S$ be the transmission, reflection, and other loss coefficients, respectively. They satisfy the equation:

\begin{equation}
T + R + S = 1. 
\end{equation}

\begin{figure}[h!]
    \centering
    \includegraphics[scale=0.08]{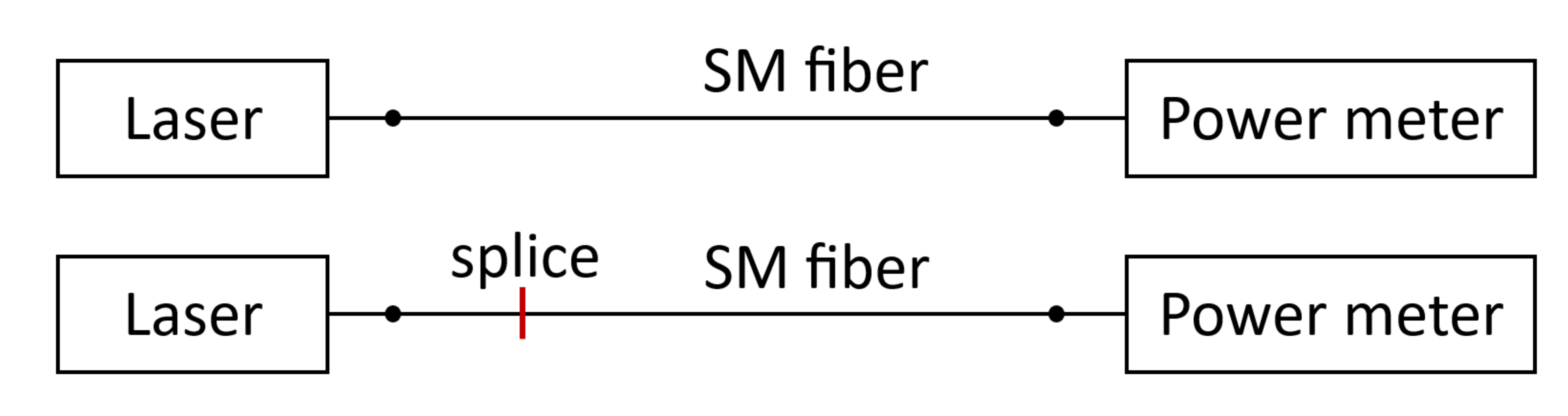}
    \caption{Experiment to measure the splice joint transmittance.}
    \label{experiment}
\end{figure}

Since the reflected signal continues to propagate through the waveguide, it is inaccessible to a potential eavesdropper. Thus, the $S$ coefficient is of interest for the calculation of the signal leakage and, therefore, it is necessary to know the values of $T$ and $R$. It is technically easier, to begin with measuring the value of $T$. The corresponding experiment scheme is shown in Fig.\,\ref{experiment}.

Each iteration is carried out in two steps. First, there is a part of a single-mode fiber, which is connected to a laser source on one side and to an optical power meter on the other side. After that, the output power data are collected for two minutes and averaged. The stability of the laser source power was also checked. The drift of the average value was about $0.01\%$ in 8 minutes, and this value decreases later as the laser gradually reaches a stable regime.

In the second step, the fiber is broken, spliced, and the average output power is measured in a similar way. This data becomes the reference for the next iteration (fiber is broken near the first joint, spliced, and the transmittance of the second splice will be measured with respect to the waveguide with one splice, etc.)
The length of the single-mode fiber section used in this experiment is no more than 10 meters, so its attenuation can be neglected.
As a result, the transmittance of each splice joint can be calculated with the formula:
\begin{equation}
T = \frac{P_{\mbox{\scriptsize{step\,2}}}}{P_{\mbox{\scriptsize{step\,1}}}}.
\end{equation}
And, consequently, the losses on Bob's side:
\begin{equation}
R + S = 1 - \frac{P_{\mbox{\scriptsize{step\,2}}}}{P_{\mbox{\scriptsize{step\,1}}}}.
\end{equation}

The results of measuring $S + R$ for the Fujikura 86S fusion splicer at SM Auto regime give the best value of 0.48\,\% and the average value of (0.65\,$\pm$\,0.28)\,\%.
It can be seen from the tabular data that the average laser source power change over 8 minutes is much less than the best loss value. Therefore, the source can be considered as being stable within one experimental iteration with good accuracy. Moreover, the random error of the average loss values exceeds the value of the average power change over 8 minutes by more than 10 times. Consequently, the contribution of laser power drift to the error of the average splice loss value is negligibly small and almost all of it consists of random splicing procedure errors (roughness of fiber cleave, splicing regime errors, etc.).

\begin{figure}[h!]
    \centering
    \includegraphics[scale=0.08]{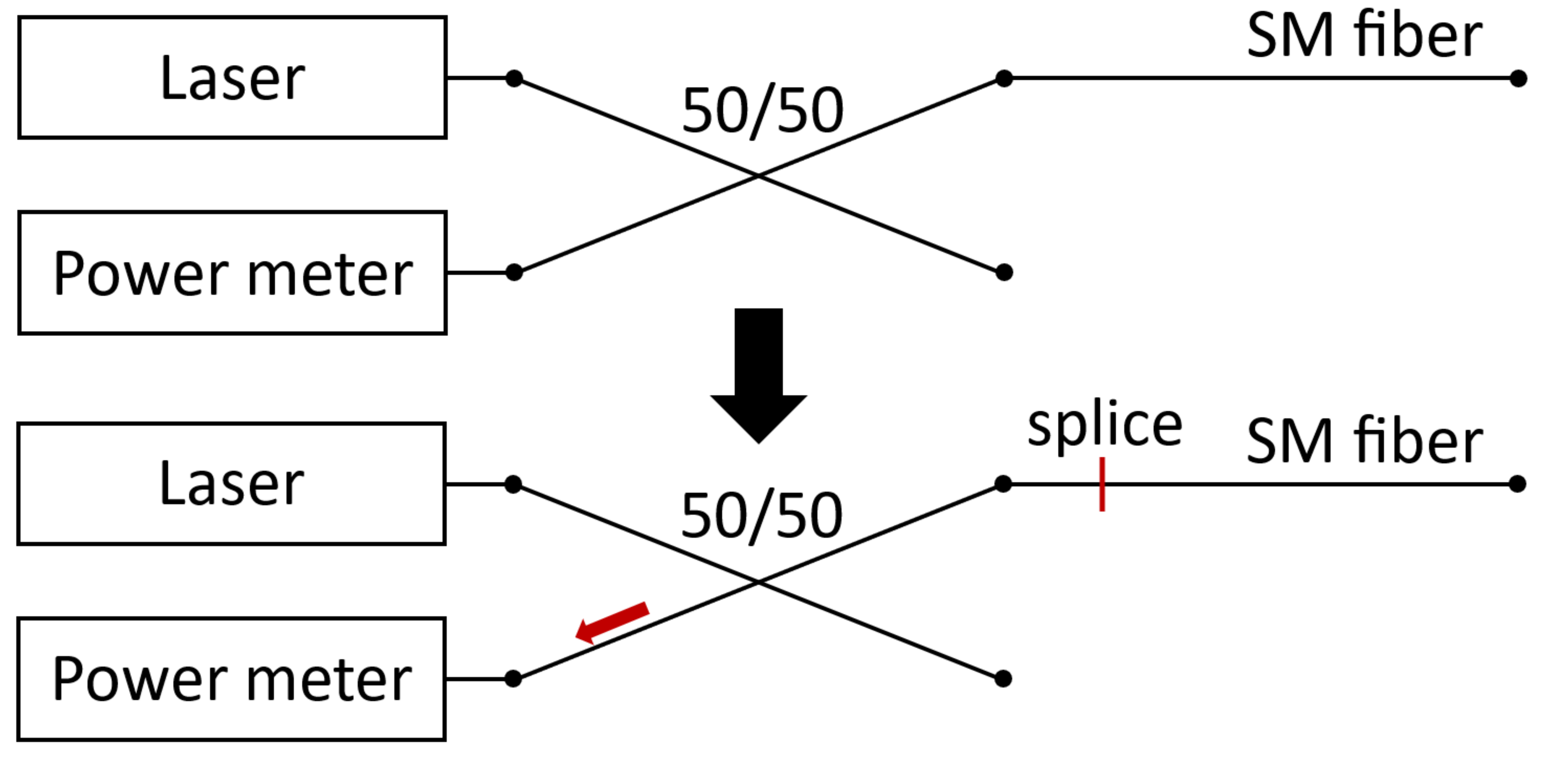}
    \caption{Reflection coefficient evaluation.}
    \label{reflection}
\end{figure}

Now, in order to calculate the $S$ value, it is necessary to investigate the reflected signal. The scheme of the corresponding experiment is shown in Fig. \ref{reflection}.

The laser signal passes through a 50/50 beamsplitter and then half of it propagates along a piece of fiber. The optical signal power is measured at the splitter output which is adjacent to the laser output. The difference between the two experimental stages is in the presence of a splice joint in the studied piece of fiber in the second stage.

In the first stage, interference with a range from 1 to 51\,nW was registered by a power meter. This signal comprises reflections from the joint between the splitter and fiber segment and reflections from the splitter and fiber ends. In the second stage, the range of interference was from 4 to 82\,nW. Therefore, with the known value of the incident power $P_{\mbox{\scriptsize{in}}}$ propagating at the first stage through the fiber segment (it is almost equal to the optical power right before the splice joint at the second stage), the reflection coefficient can be estimated by the formula:
\begin{equation}
R \leq \frac{2(P_{\mbox{\scriptsize{max}}}^{(2)} - P_{\mbox{\scriptsize{max}}}^{(1)})}{P_{\mbox{\scriptsize{in}}}},
\end{equation}
where $P_{\mbox{\scriptsize{max}}}^{(j)}$ is the maximum value of reflected signal power measured at the $j$-th stage.
For $P_{\mbox{\scriptsize{in}}} = 10$\,mW $$ R \leq 6.2 \times 10^{-6} < 10^{-5}. $$

Thus, even the upper estimate of $R$ is much less than the sum $S + R$, which allows us to conclude that almost all the power lost for Bob's side does not remain in the reflected signal. Or, in a mathematical form, $S + R \approx S$, $S \approx 1 - T$.
As a result, at present, even an automatic program on the common commercial splicing device allows achieving a loss value less than 1$\%$ on average and around 0.5$\%$ in good cases.

\subsection*{Emitted power}
\begin{figure}[h!]
    \centering
    \includegraphics[width=0.84\linewidth]{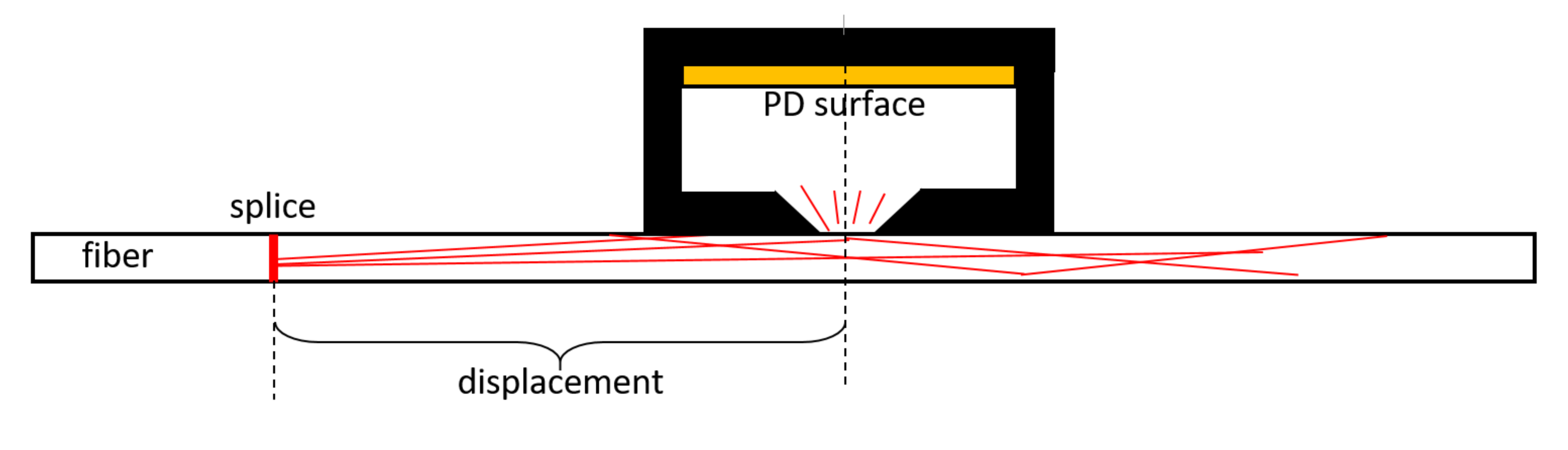}
    \caption{Measurement scheme.}
    \label{measur_scheme}
\end{figure}

Nevertheless, not all optical power lost on the splice joint is scattered into the external environment and can be collected by an eavesdropper eventually. For quantitative research of the emitted power fraction, the experiment, a schematic picture of which is given in Fig.\,\ref{measur_scheme}, was carried out for the three splices with different loss values $S$. 

In this experiment, a photodetector with a special cap which increases the spatial resolution is moved along the fiber and the optical power is recorded during the transmission of the laser radiation through the splice joint.
The experimental data represent the emitted power dependence on the detector's center displacement relative to the splice. The graph example is given below in Fig.\,\ref{emitted_power}.

\begin{figure}[h!]
    \centering
    \includegraphics[scale=0.12]{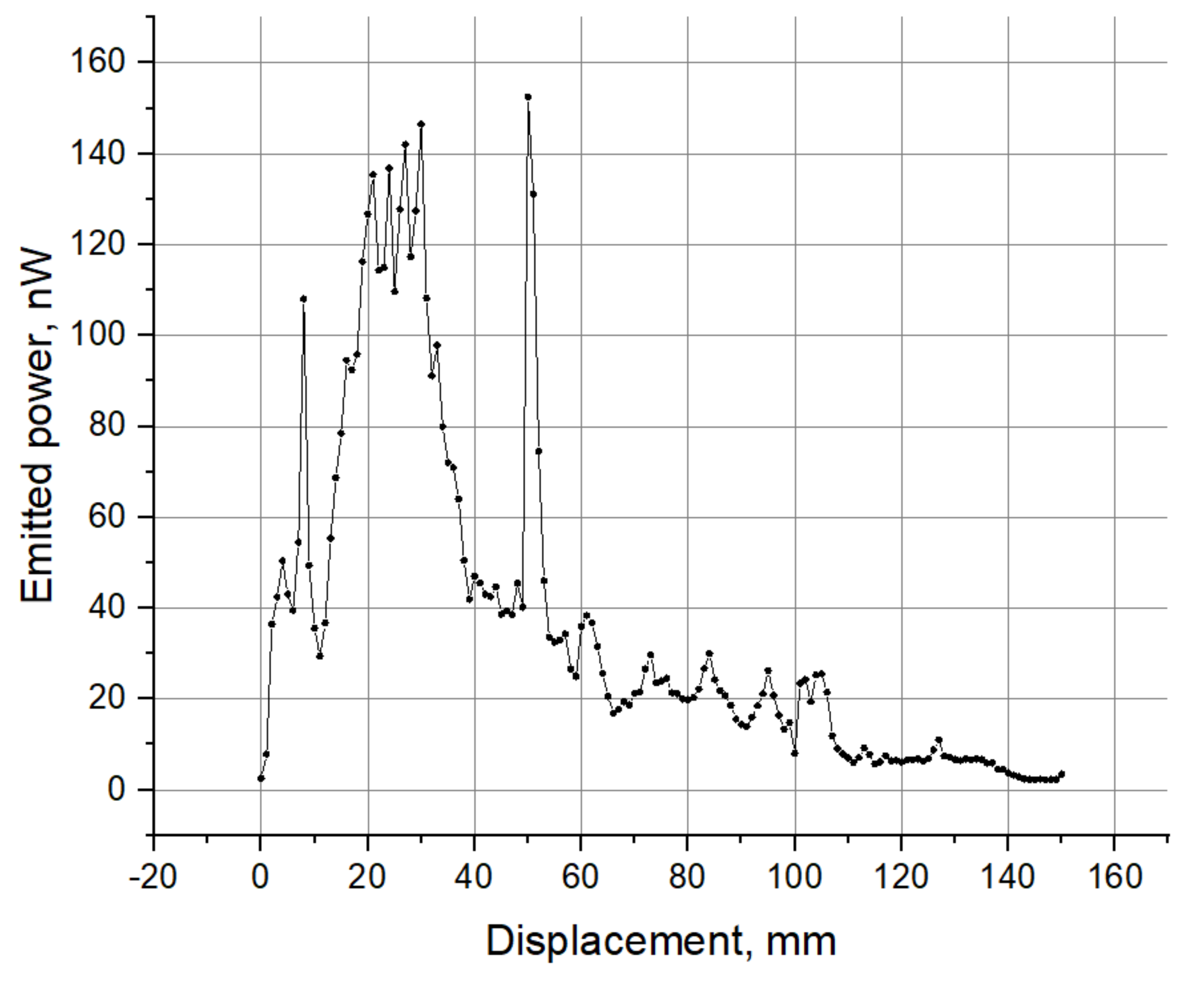}
    \caption{Emitted power distribution for splice with $S = 0.55\%$.}
    \label{emitted_power}
\end{figure}

By integrating the function plotted in this graph and taking into account that at each point only a part of the radiation is collected by photodetector, it is possible to calculate the integral value of the emitted optical power $P_e$. 
As a result, it turns out to be around 1/4 of the $S$ value on average, as presented in Table \ref{losses_table:2}.

\begin{table}[h!]
\centering
\begin{tabular}{|rrrr|r|}
\hline
\multicolumn{1}{|r|}{№ of splice} & \multicolumn{1}{r|}{$P_\text{in}$, mW} & \multicolumn{1}{r|}{$S$, \%} & $\Tilde{S}  = P_e/P_\text{in}$, \% & $\Tilde{S}/S$   \\ \hline
\multicolumn{1}{|r|}{1}        & \multicolumn{1}{r|}{18.38}       & \multicolumn{1}{l|}{0.65}  & 0.162                & 0.25         \\ \hline
\multicolumn{1}{|r|}{2}        & \multicolumn{1}{r|}{18.26}       & \multicolumn{1}{l|}{0.33}  & 0.095                & 0.29         \\ \hline
\multicolumn{1}{|r|}{3}        & \multicolumn{1}{r|}{18.20}       & \multicolumn{1}{l|}{0.55}  & 0.132                & 0.24         \\ \hline
\multicolumn{4}{|l|}{Average}                                                                                         & 0.26 $\pm$ 0.03 \\ \hline
\end{tabular}
\caption{}
\label{losses_table:2}
\end{table}

\begin{figure}[h!]
    \centering
    \includegraphics[width=0.84\linewidth]{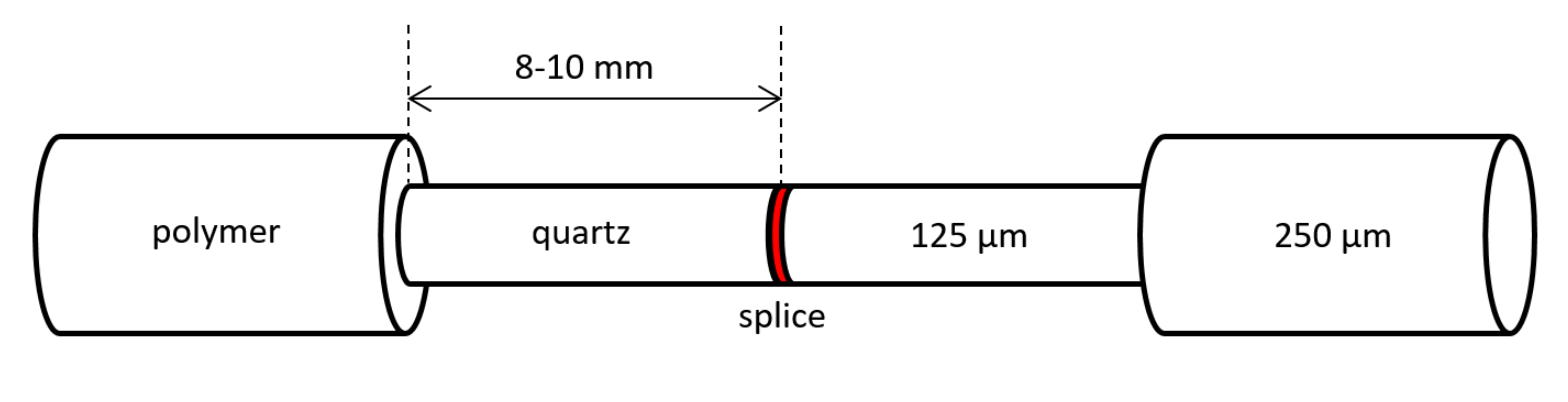}
    \caption{Schematic representation of a SM fiber splice joint without protection sleeve.} 
    \label{splice_joint}
\end{figure}

It is worth noticing here, that all the measurements were done on splices without any additional protection, the example of which is shown below in the Figure \ref{splice_joint}. Since in real conditions splices are usually protected with special sleeves, we also have investigated its influence on the integral emitted power. As a result, the reduction of 10-15\% was observed and by taking this effect into account, one can derive that in real conditions external environment losses are $0.23\pm0.05$ of the total loss value on the given splice joint.

\subsection*{Leaky modes}

Since the splice connects two optical waveguides, then by applying the leaky modes theory\,\cite{losses} to the ITU-T G.652.D single-mode fiber, the values of the propagation angle $\theta_z$ far from the core, e.g. near the quartz cladding's edge can be obtained for the first few $HE_{l+1, m}$ modes of the higher order (Table \ref{table_losses:3}).

\begin{table}[h!]
\centering
\begin{tabular}{|l|l|}
\hline
$l,\,m\quad$ & $ \tan{\theta_z}$ \\
\hline
1, 1 & $9.88\times10^{-3}$ \\
\hline
2, 1 & $8.33\times10^{-2}$ \\
\hline
3, 1 & 0.14 \\ \hline
\end{tabular}
\caption{Results of leaky modes theory for the ITU-T G.652.D fiber.}
\label{table_losses:3}
\end{table}

\begin{figure}[h!]
    \centering
    \includegraphics[width=0.84\linewidth]{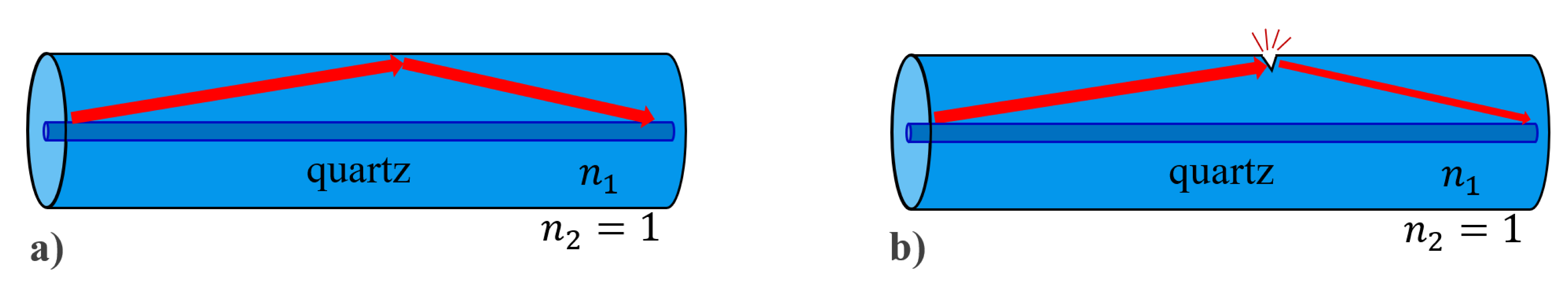}
    \caption{Schematic illustration of the: a) leaky mode reflection, b) leaky mode scattering into the external environment.}
    \label{leaky_mode}
\end{figure}

All those values correspond to the case of total internal reflection from the quartz-air or polymer(acrylate)-air boundaries. However, emitted power was observed at distances of several centimeters after the splice joint during the experiments mentioned above, which means that this radiation could be scattered by some boundary defects. The illustration of this process is shown in Fig.\,\ref{leaky_mode}.

\begin{figure}[h!]
    \centering
    \includegraphics[width=0.84\linewidth]{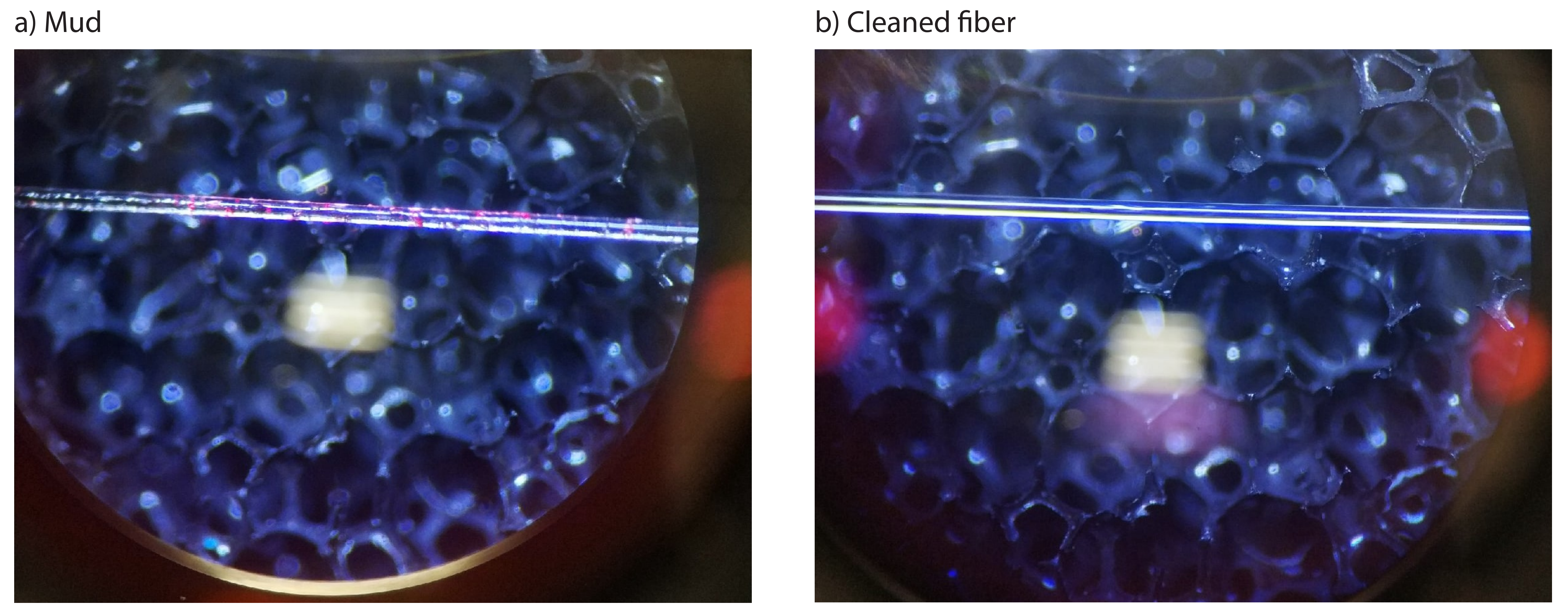}
    \caption{Observation of a 125 µm cladding under a microscope.}
    \label{fig:test2}
\end{figure}

To prove this assumption, a piece of the fiber after the splice joint was visually observed with a microscope when transmitting red laser radiation through it (Fig.\,\ref{fig:test2}).

One can see that there is a decrease in red light emission due to the good cleaning of quartz because pieces of mud, like cracks, are boundary defects too and consequently can lead to the escape of leaky modes from the waveguide. The integral value of emitted power, measured on the cleaned fiber right after the splice joint, was 5.74\,nW or

\begin{center}
$3\times10^{-7}$ of the incident power ($P_{in} = 18$ mW).    
\end{center}

Thus, with the polymer removal and good cleaning of the quartz, a significant decrease in the emitted power can be achieved. A further task is the implementation of this technique in real-life conditions where it is necessary to use fiber protection, which can potentially change the refractive indices difference and lead to the emission of the leaky modes into the external environment again.

\bigskip\section*{NOTE 3. Used equipment}
On the Alice side, the TLX1 Thorlabs laser is used to create laser radiation. Next, the iXblue MPZ-LN-10 phase modulator is in use for phase randomization of the generated laser radiation. Finally, the iXblue MXAN-LN-10 amplitude modulator is applied for intensity modulation. The signal coming from the FPGA to the amplitude modulator is amplified using the iXblue DR-VE-10-MO amplifier.

On the Bob side, the Menlo Systems FPD610-FC-NIR detector is used to detect the optical signal. Then the analog signal from the detector is digitized by the Spectrum Devices M4i.2210-x8 ADC.

\bigskip\section*{NOTE 4. Temperature dependence}\label{temp_dep_section}
\begin{figure}[h!]
    \centering
    \includegraphics[width = 0.6\linewidth]{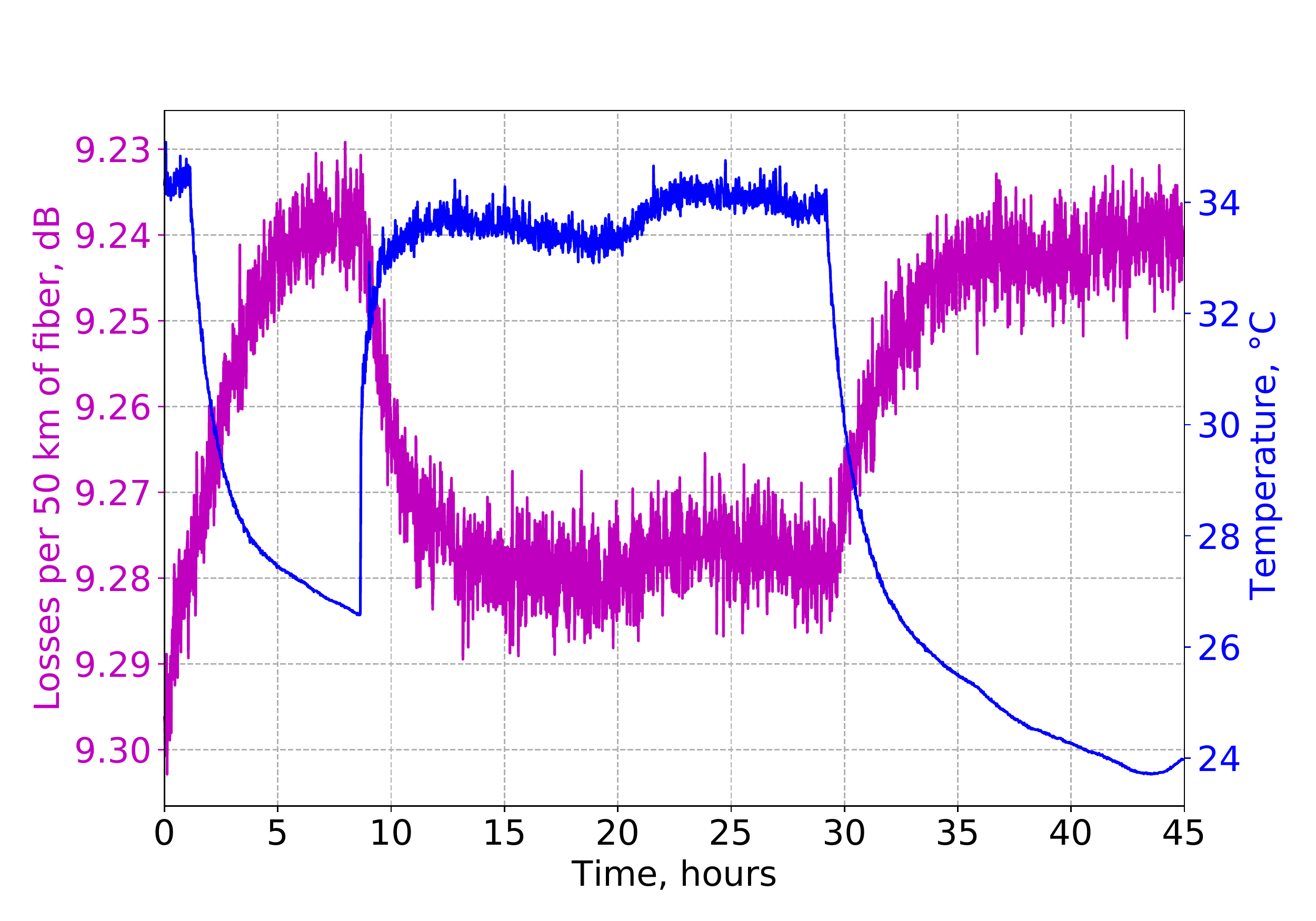}    
    \caption{The time dependencies of the losses in the optical fiber and the fiber temperature. The temperature change is shown by the blue curve, the corresponding loss coefficient behavior is presented by the magenta one.}
    \label{OTDR_loss_check}
\end{figure}

The optical loss control in the communication line enables not only determining the magnitude of the signal diverted by the eavesdropper but also the natural changes in the optical fiber transparency. One of the reasons for the transparency coefficient change is the temperature variations. Using the OTDR, we measure the losses magnitude change at the coil containing the 50\,km of the fiber line upon the coil heating, see Fig.\,\ref{OTDR_loss_check}. The obtained reflectogram allows us to determine the loss magnitude of the whole segment of the optical fiber line from the slope of the log-log plot.   

\bigskip\section*{NOTE 5. The narrowband filter}\label{narrow_filter_section}
\begin{figure}[t]
    \centering
    \includegraphics[scale=0.3]{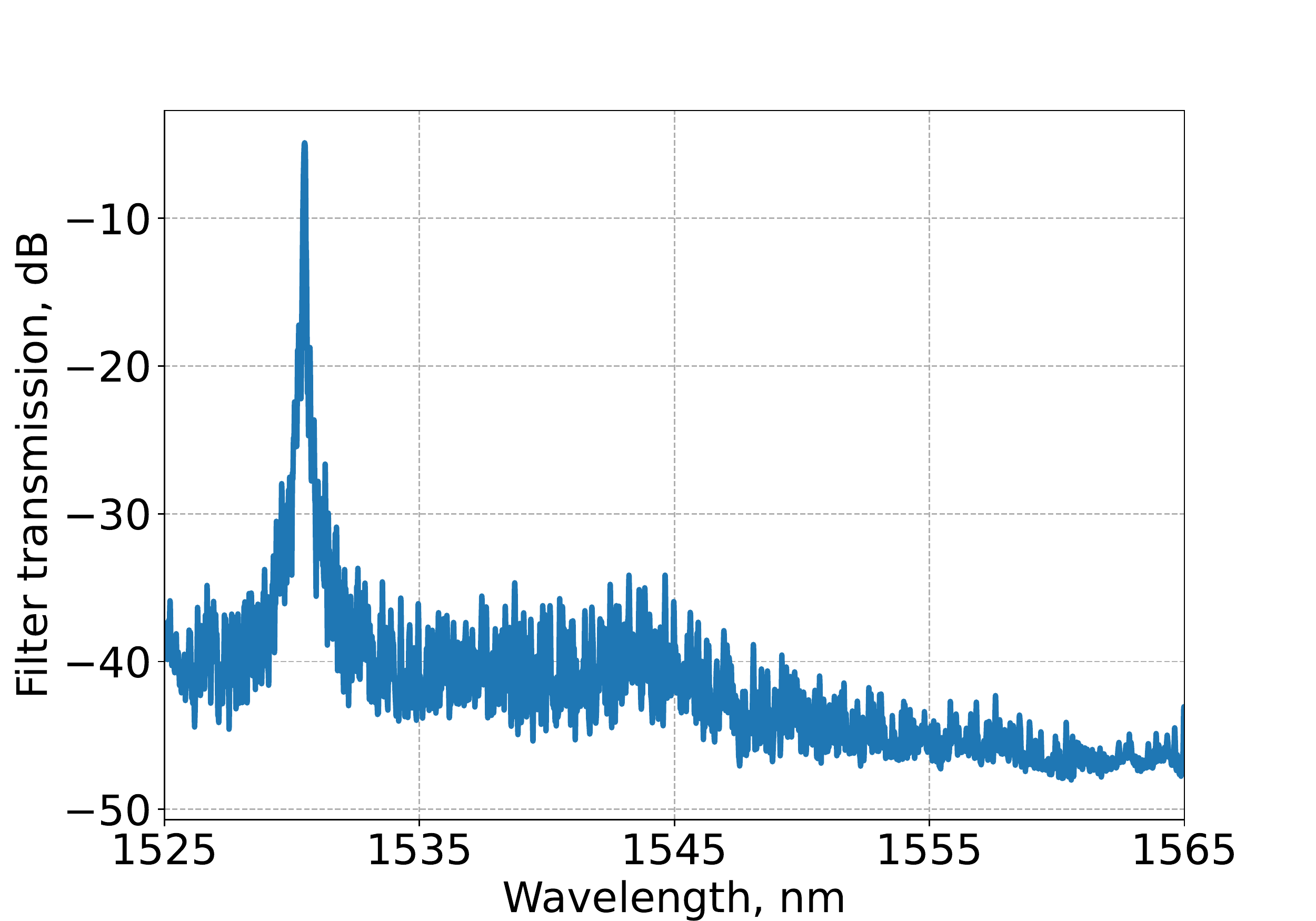}
    \caption{The dumping spectrum of the narrowband filter for the dumping of 3\,dB.}
    \label{8_5GHz_filter}
\end{figure}

To reduce noise in the optical line it is necessary to suppress parasitic modes that do not carry information. For this purpose Terra Quantum AG has developed a special narrowband filter. Here in Fig.\,\ref{8_5GHz_filter} we present the transmission spectrum of the narrowband filter having a bandwidth of 8.5\,GHz with 3\,dB dumping.

\bigskip\section*{NOTE 6. Classical post-processing}
Standard post-processing comprises three main stages: postselection, error correction and privacy amplification -- each of the stages is reflected in Fig.\,\ref{protocol_block_scheme}. We discuss the first stage in detail in the Sec.\,\ref{protocol_section}, while the specifics of the later are addressed below.

\subsection*{Error correction}\label{SI_error_correction}
For the purpose of correcting the resulting error, we utilize LDPC (linear-density-parity-check) code\,\cite{MacKay} practical realization.
Usually, the input for such codes are probabilities of zero and one at each position of the raw key.
Here, we deliver to LDPC only bit error rate $p_{\mbox{\scriptsize{err}}}$ averaged over the whole sequence.
Bob estimates BER according to the their measurement results, see Fig.\,\ref{distribution_1}(a).
During error correction legitimate users disclose some information about the key to Eve depending on the syndrome's length.
An accurate amount of the disclosed information is $f\cdot h_2\left( p_{\mbox{\scriptsize{err}}} \right)$, where $f$ is the effectiveness.

\subsection*{Privacy Amplification}\label{SI_privacy_amplif}

To increase the secrecy of the corrected key, it is necessary to compress it to such a size that Eve will have practically no information about it. Let $A$ and $B$ be the sets of adjusted sequences of length $m$ and final keys of length $n$, respectively. Then, to compress an element from the set $A$ into an element of the set $B$, we use the Toeplitz matrix $T$ of size $n \times m$, whose elements are random binary numbers. The set of Toeplitz matrices $T: A \longrightarrow B$ form the class of 2-universal hash functions $H \ni T$. According to the definition of Carter and Wegman\,\cite{u2_hash} a class $H$ of hash functions from $A$ to $B$ is called 2-universal if there are at most $\frac{|H|}{|B|}$ hash functions $h \in H$ such that $h(a_1) = h(a_2)$ for any $a_1 \ne a_2 \in A$. In paper\,\cite{u2_toeplitz}, the 2-universality of Boolean matrices was shown, of which binary Toeplitz matrices are a special case.
For $|H| = 2^{m+n-1}$ and $|B| = 2^n$ collision probability $P_{\mbox{\scriptsize{collision}}} = \frac{|H|/|B|}{|H|} = \frac{1}{|B|} = 2^{-n} $. Therefore, the longer the final key, the less chance of a collision. In real conditions, we compress keys in portions of at least a thousand bits.

\bigskip\section*{NOTE 7. Current state of reflectometry}\label{loss_section}
\begin{figure}[h!]
    \centering
    \includegraphics[width = \linewidth]{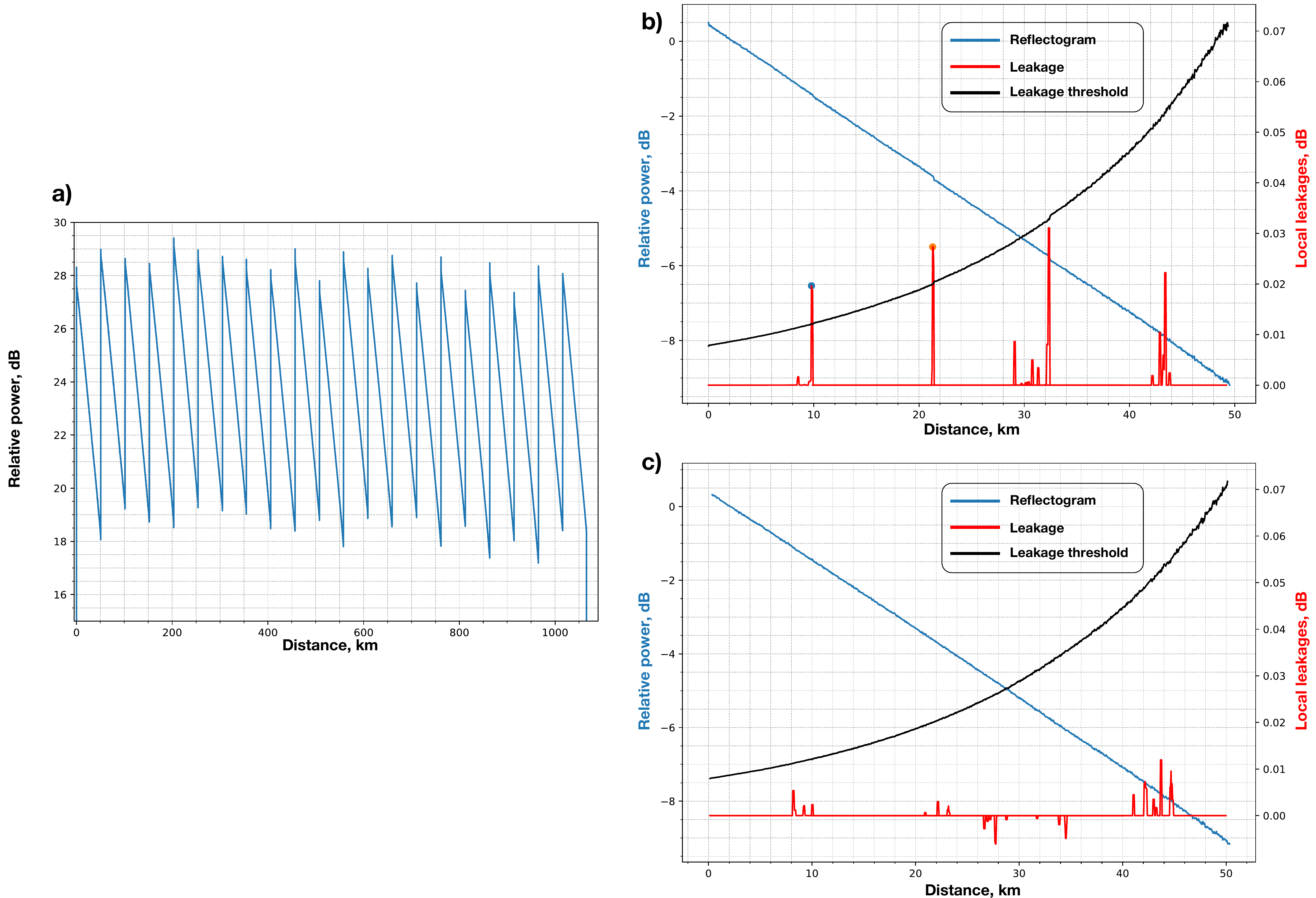}
    \caption{\textbf{Current status of the matter of reflectometry. a)} Trace of the entire line obtained with a TQ-made OTDR. Current measurements are made on the new line. \textbf{b)} Reflectogram of a separate section after amplifier 20. In this section 4 leaks were introduced, which are shown by red peaks. Also the threshold of leakage to the interceptor is shown in black on the graph. \textbf{c)} Reflectogram of a separate section after 19 amplifier without introduced additional losses.}
    \label{Refl_example}
\end{figure}

At the moment, the probing pulse duration is 300\,ns, the number of averages is 5000, and the laser power is 20\,mW. In Fig.\ref{Refl_example} we demonstrate the current state of the matter. 

\end{document}